\documentclass[a4paper,11pt]{article}
\pdfoutput=1 % if your are submitting a pdflatex (i.e. if you have
\usepackage{jcappub}
\usepackage{graphicx}
\usepackage{color}
\usepackage[dvipsnames]{xcolor}
\usepackage{xspace}
\usepackage{multirow}
\usepackage{cleveref}		% Automatic eq. fig., etc.
\usepackage[normalem]{ulem}
\usepackage{makecell}
\usepackage{placeins}
\usepackage{capt-of}

\newcommand{\secref}[1]{Sec.~\ref{#1}}
\newcommand{\fulleqref}[1]{Eq.~\eqref{#1}}

\newcommand{\tquote}[1]{``#1''}

\newcommand{\dnew}{\par\vspace{0.5\baselineskip}\mbox{}}
\newcommand{\dnewnoindent}{\par\vspace*{0.5\baselineskip}\noindent\mbox{}}
\providecommand{\abs}[1]{\ensuremath{\lvert#1\rvert}}

\providecommand{\SD}{SD\xspace}

\providecommand{\lcdm}{$\Lambda$CDM\xspace}
\DeclareMathAlphabet\mathbfcal{OMS}{cmsy}{b}{n}

\usepackage{csquotes}
\usepackage[all]{nowidow}

\usepackage{titlesec}
\titlespacing\section{0pt}{20pt plus 4pt minus 2pt}{12pt plus 2pt minus 2pt}
\titlespacing\subsection{0pt}{14pt plus 4pt minus 2pt}{10pt plus 2pt minus 2pt}
\titlespacing\subsubsection{0pt}{12pt plus 4pt minus 2pt}{8pt plus 2pt minus 2pt}

\author[1]{Nils Sch\"{o}neberg,}
\author[2]{Matteo Lucca,}
\author[2]{and Deanna C. Hooper}

\affiliation[1]{Institute for Theoretical Particle Physics and Cosmology (TTK), \\ RWTH Aachen University, D-52056 Aachen, Germany.}
\affiliation[2]{Service de Physique Th\'{e}orique, \\Universit\'{e} Libre de Bruxelles, C.P. 225, B-1050 Brussels, Belgium}

\emailAdd{schoeneberg@physik.rwth-aachen.de}
\emailAdd{matteo.lucca@ulb.be}
\emailAdd{deanna.hooper@ulb.be}

\title{Constraining the inflationary potential with spectral distortions}

\abstract{
Measuring spectral distortions (SDs) of the cosmic microwave background (CMB) will provide new constraints on previously unexplored scales of the primordial power spectrum, allowing us to extend the probed parameter space by several orders of magnitude in $k$-space, which could have significant implications in the context of primordial black holes and gravitational waves, among others. Here we discuss how various models of inflation can be tightly constrained by the combination of current and future CMB SD and anisotropy experiments. In particular, we investigate the constraining power of SD experiments such as FIRAS, PIXIE, and PRISM in conjunction with CMB anisotropy probes such as Planck or CMB-S4 plus LiteBIRD. Building on the latest version of the Boltzmann solver \textsc{class} (v3.0), here we also consistently marginalize over the possible galactic and extra-galactic foregrounds for the SD missions. With this numerical setup, we are able to realistically forecast the improvements that the increased lever-arm provided by the addition of the various SD missions will bring for several combinations of the aforementioned experiments. As a result, in all considered models we observe that SDs provide a highly significant tightening of the constraints by up to 640\%, and increase the figure of merit up to a factor of around 1600.}

\begin{document}

\hfill{\small TTK-20-33}\\
\vspace{-0.5 cm}
\hfill{\small ULB-TH/20-14}

\vspace*{-2\baselineskip}\vspace*{0.5cm}

\maketitle

\section{Introduction}
Despite its conceptual and mathematical simplicity, slow-roll inflation \cite{BROUT197878, STAROBINSKY198099, Kazanas1980, Sato1981, LINDE1982389, LINDE1983177, Albrecht1982, Stewart:1993bc, Liddle:1994dx} has been shown to be an incredibly successful model \cite{Planck:2013jfk, Ade:2015lrj, Ade:2015tva, Akrami2018PlanckX}. In fact, the minimal idea of a scalar field slow-rolling in a very flat potential is enough to consistently solve several issues, such as the flatness, horizon, entropy, and monopole problems, as well as the origin of adiabatic primordial perturbations, otherwise unexplained in the standard Big Bang cosmology (see e.g., \cite{Linde:2005ht, Baumann:2009ds} for comprehensive reviews).
\dnew
One direct consequence of the flatness of the inflationary potential is the almost perfect scale-invariance of the primordial power spectrum (PPS) of the metric perturbations. This property of the PPS is assumed to be true at all scales and, therefore, affects several cosmological observables, such as the overall amplitude and the tilt of the Cosmic Microwave Background (CMB) anisotropy power spectra, as well as the shape of the CMB spectral distortions (SDs) \cite{Zeldovich1969Interaction, Sunyaev1970Interaction, Zeldovich1972Influence, Illarionov1975ComptonizationI, Danese1982Double, Burigana1991Formation, Hu1993ThermalizationI, Chluba2011Evolution, Fu:2020wkq} (for recent reviews see e.g., \cite{Lucca2019Synergy}, in particular Sec. 2.4.3 therein, as well as \cite{Kogut2019CMB, Chluba2019Voyage}).
\dnew
However, this assumed scale-invariance needs to be tested across as many different scales as possible. Although many decades of wave numbers in Fourier space have already been constrained using CMB anisotropies observations \cite{Planck:2013jfk, Ade:2015lrj, Ade:2015tva, Akrami2018PlanckX} -- together with Baryon Acoustic Oscillation (BAO) data -- a large part of parameter space is still unexplored, as shown for instance in \cite{Byrnes2018Steepest} (see e.g., Fig. 8 therein). In this regard, SDs are an ideal complementary probe to CMB anisotropies. In fact, as they are produced in the epoch previous to recombination by the gradual inefficiency of number changing processes (like Bremsstrahlung and double Compton scattering) and scattering processes (such as Compton scattering), SDs allow to probe completely different scales than those constrained by anisotropy observations. Explicitly, while CMB anisotropies cover scales between $10^{-4}$ Mpc$^{-1}$ and $0.5$ Mpc$^{-1}$ (see e.g., Fig. 20 of \cite{Akrami2018PlanckX}), SDs are mainly influenced by scales between $1$ Mpc$^{-1}$ and $10^4$ Mpc$^{-1}$ (see e.g., Fig. 9 of \cite{Fu:2020wkq}).
\dnew
Exploiting this complementarity, in this work we make use of the numerical tools developed in \cite{Lucca2019Synergy, Fu:2020wkq} (incorporated in the public codes \textsc{class}~\cite{Lesgourgues2011Cosmic, Blas2011Cosmic} and \textsc{MontePython}~\cite{Audren2013Conservative, Brinckmann2018MontePython}) to forecast possible future constraints on the PPS over more than eight orders of magnitude in Fourier space. 
In particular, here we consider several different parametrizations of the PPS, namely higher-order expansions of the inflationary potential or features such as steps, kinks, and two-component models. 
While higher order expansions of the PPS play an important role when investigating model-independent scenarios beyond the simple slow-roll approximation (see e.g., \cite{Grivell:1999wc, Lesgourgues:2007gp, PhysRevD.82.043513, Vazquez:2012ux} and references therein, as well as \cite{Handley:2019fll} for a more recent discussion), features in the PPS have been employed in the literature in the context of hybrid inflation scenarios \cite{Bugaev:2011qt, Hazra:2014jka, Hazra:2014goa, Hazra:2016fkm}, primordial gravitational waves (GWs) \cite{Hotchkiss:2011gz}, and primordial black holes (PBHs) \cite{Byrnes2018Steepest} among many others (see \cite{Chluba:2015bqa} and references therein for an extensive review). Furthermore, more in general, improved constraints on the inflationary epoch from CMB SDs have also been shown to have interesting applications for very modern topics in cosmology such as the primordial GW background \cite{Kite2020Bridging} and the Hubble tension \cite{Lucca:2020fgp}.
\dnew
Moreover, in order to present more realistic constraints, we improve on the work presented in~\cite{Lucca2019Synergy, Fu:2020wkq} by including a marginalization over all galactic and extra-galactic foregrounds relevant for the considered SD missions. To do so, we base our implementation on the extensive analysis proposed by the Planck collaboration \cite{Adam2015PlanckX, Akrami2018PlanckIV} and \cite{Abitbol2017Prospects}.
\dnew
This paper is organized as follows. We begin in Sec.~\ref{sec:th_inf} with a brief review of the theory of inflation, with a special focus on parametrizing the inflationary potential. In Sec.~\ref{sec:th_imprints} we discuss the impact of inflation on cosmological observables such as SDs of the CMB, and provide a detailed description of which foregrounds and secondary effects need to be considered. In Sec.~\ref{sec:num} we discuss the numerical setup used for our analysis, while in Sec.~\ref{sec:res} we present our main results. We then conclude in Sec.~\ref{sec:con} with a summary, and provide further discussion on the generation of anisotropies in App.~\ref{app:th_CMB}.

\section{Inflation}\label{sec:th_inf}

In the following section we review the main and most general features of inflation. The overall notation and most of the equations are based on \cite{Baumann:2009ds} and references therein, while a comprehensive overview can be found e.g., in \cite{Lesgourgues2006Inflationary}.
\dnew
In curved space-time, a single-component field $\varphi$ evolving in a potential $V(\varphi)$ can be described by the
action
\begin{align}\label{eq:action}
	S=\int d^{4} x \sqrt{|g|}\left(\frac{R}{2}+\frac{1}{2} \partial_{\mu} \varphi \partial^{\mu} \varphi-V(\varphi)\right)~,
\end{align}
where $g$ is the modulus of the metric tensor $g_{\mu\nu}$ and $R$ is the Ricci scalar. Note that we adopt units of $8 \pi \mathcal{G}=c=1$, where $\mathcal{G}$ is the gravitational constant.
\dnew
Varying the action of Eq. \eqref{eq:action} with respect to the metric leads to the definition of the energy-momentum tensor and corresponding Einstein equations as
\begin{equation}
T_{\mu \nu}=\partial_{\mu} \varphi \partial_{\nu} \varphi-\mathcal{L}_{\varphi} g_{\mu \nu}=G_{\mu\nu}~.
\end{equation}
From the $(00)$ and the $(ii)$ components of this equation one obtains the well-known Friedmann and acceleration equations for the homogeneous background, which are respectively
\begin{align}
	& H^{2} \equiv \left(\frac{\dot{a}}{a}\right)^2=\frac{1}{3} \rho~, \label{eq:Fr1} \\
	& \frac{\ddot{a}}{a}=-\frac{1}{6}(\rho+3 p)~.  \label{eq:Fr2}
\end{align}
Here $H=\dot{a}/a$ is the Hubble rate, while $\rho$ and $p$ correspond respectively to the total energy density and pressure of the system, which in the case of the action \eqref{eq:action} and assuming a homogeneous field are given by
\begin{align}\label{eq:rho_p}
	\rho=\frac{1}{2} \dot{\varphi}^{2}+V \quad \text{and} \quad p=\frac{1}{2} \dot{\varphi}^{2}-V\,.
\end{align}
Combining Eqs. \eqref{eq:Fr1} and \eqref{eq:Fr2}, it can be shown that the energy density evolves according to the conservation equation
\begin{align}\label{eq:cons_rho}
	\dot{\rho}=-3 H\left(\rho+p\right)\,.
\end{align}
Defining inflation as an accelerated stage of expansion leads to the condition
\begin{align}\label{eq:def_inflation}
	\frac{\ddot{a}}{a} = \dot{H} + H^2 = -\frac{1}{6} (\rho + 3 p) > 0~.
\end{align}
In terms of $\epsilon_H = -\dot{H}/H^2 = \dot{\varphi}^2/H^2$ this definition is equivalent to $\epsilon_H<1$.\footnote{This is also how the inflation condition is implemented in the \textsc{class} code, even for beyond slow-roll scenarios.} 
\dnew
At the level of the homogeneous background, the equation of motion of the field $\varphi$ is simply the Klein-Gordon equation
\begin{align}\label{eq:kleingordon}
	\frac{1}{\sqrt{|g|}} \partial_{\mu}\left[\sqrt{|g|} \partial^{\mu} \varphi\right]+\partial_{\varphi} V=\ddot{\varphi}+3 H \dot{\varphi}+\partial_{\varphi} V=0\,,
\end{align}
and can be obtained by varying the action Eq. \eqref{eq:action} with respect to the field $\varphi$. The perturbed evolution of the field can be phrased in terms of the \enquote{comoving curvature perturbation}
\begin{align}
	\mathcal{R} = \Psi - \frac{H}{\rho+p} \delta q = \Psi + \frac{H}{\dot{\varphi}} \delta \varphi~,
\end{align}
where $\delta T^0_i = \partial_i\delta q$, $\Psi$ is the Bardeen potential, and the second equality holds for single-field inflation, where $\delta T^0_i = -\dot{\varphi} \partial_i \delta \varphi$. In this case, the description through $\mathcal{R}$ is equivalent to that through $\delta \varphi$. The initial conditions of the adiabatic perturbations throughout the later evolution of the universe can be related to the comoving curvature perturbation on super-Hubble scales.
\newpage 
\dnewnoindent
We can then derive the Mukhanov equation (see e.g., App. B.2 of \cite{Baumann:2009ds})
\begin{align}\label{eq:mukhanov}
	v'' + \left(k^2 - \frac{z''}{z}\right) v = 0~,
\end{align}
with $v = z \mathcal{R}$ and $z = a \dot{\varphi}/H$. After obtaining the solution of $v(k,a)$ we make use of the fact the curvature perturbation freezes out on super-Hubble scales ($aH \gg k$) to define the primordial power spectrum
$P_\mathcal{R}(k)$ as
\begin{align}
	 (2\pi^3) \delta(\mathbf k + \mathbf k') P_\mathcal{R}(k) = \langle \mathcal{R}(\mathbf k) \mathcal{R}^*(\mathbf k') \rangle = \abs{v}^2/z^2~,
\end{align}
with the frozen-out super-Hubble limit of $v$.\footnote{In the \textsc{class} code, the inflationary equations are evolved until the relative variation of $v(k,a)$ falls below arbitrary small value.} Finally, as often done in the literature, we introduce the dimensionless PPS
\begin{align} \label{eq:dim_pps}
	\mathcal{P}_\mathcal{R}(k) = \frac{k^3}{2\pi^2} P_\mathcal{R}(k)~.
\end{align}

\subsection{Inflationary potential}\label{ssec:potential}
In principle, the theoretical prescription described above does not require any slow-roll approximation. In fact, results can be found for any arbitrary potential $V(\varphi)$. For instance, under the assumption that the potential can be Taylor-expanded around $\varphi=0$, it is possible to reconstruct the total potential with a finite Taylor approximation up until  order $\varphi^N$, thus obtaining
\begin{align} \label{eq:Vexpansion}
V(\varphi) = V_0 + V_1 \varphi + \frac{V_2}{2!} \varphi^2 + \frac{V_3}{3!} \varphi^3 + ... + \frac{V_N}{N!} \varphi^N = \sum\limits_{n=0}^N \frac{V_n}{n!} \varphi^n~.
\end{align}
\dnewnoindent
For such a potential, we can also introduce the potential slow-roll (PSR) parameters
\begin{align}\label{eq:SR_params}
\epsilon_{V} =\frac{1}{2} \left(\frac{V_1}{V_0}\right)^{2}\,, \quad \eta_{V} = \frac{V_2}{V_0}\,, \quad \xi_{V} = \frac{V_1 V_3}{V_0^2}\,, \quad \omega_{V} = \frac{V_1^2 V_4}{V_0^3}\,,\quad s^{(n)}_V = \frac{V_1^{n-1}V_{n+1}}{V_0^n}~.
\end{align}
These, as will be discussed further in Section \ref{ssec:th_slow_roll}, are subject to the additional conditions $\epsilon_V \ll 1$ and $\abs{\eta_V} \ll 1$ when assuming slow-roll. However, in full generality these are not necessarily related to slow-roll, and are instead equivalent to such a Taylor expansion.
\dnew
The assumption of a Taylor expansion existing can either be imposed until the end of inflation, or only in the observable window, thus making no assumptions about the end or duration of inflation. Here we assume the latter, more conservative, case (called \texttt{inflation\_V} in \textsc{class}). With this assumption, we can use the inflationary module of \textsc{class} to compute the primordial spectrum with a full integration of the Fourier mode evolution. For more technical aspects on this assumption, we refer to~\cite{Lesgourgues:2007gp}.

\subsection{Slow-roll}\label{ssec:th_slow_roll}
Within this work we also consider the more commonly-used approximation of slow-roll inflation, although we only impose this assumption in Sec.~\ref{ssec:results_SR}. 
From Eqs. \eqref{eq:rho_p} and \eqref{eq:def_inflation} it follows that $\dot{\varphi}^{2} < V$ at any time during inflation. In order to be in slow-roll inflation, an expansion very close to the de Sitter limit is needed, thus requiring $\dot{H} \ll H^2$ and equivalently $\epsilon_H \ll 1$. In this limit, one obtains $\dot{\varphi}^2 \ll V$ and $\rho \approx V \approx -p$. A second slow-roll condition can be derived by imposing the first slow-roll condition to hold for an extended period of time, thus requiring $\abs{\ddot{\varphi}} \ll \abs{\partial_\varphi V}$, which can equivalently be stated in terms of the potential slow-roll parameters as $\epsilon_{V}\ll1$ and $|\eta_{V}|\ll1$.\footnote{In slow-roll, one also has $\epsilon_H \approx \epsilon_V$.}
\dnew
For such potentials it is possible to derive (using the quasi-de Sitter solutions of Eq.~\eqref{eq:mukhanov}) the following equation
\begin{align}
	\mathcal{P}_\mathcal{R}(k) \approx \frac{H^2}{8\pi^2}\frac{1}{\epsilon_H} \bigg\rvert_{k=aH} \approx  \frac{V}{24\pi^2}\frac{1}{\epsilon_V} \bigg\rvert_{k=aH}\,.
\end{align}
We can now express the Taylor expansion of the power spectrum around some arbitrary pivot scale $k_*$ in terms of the potential slow-roll parameters as 
\begin{align}\label{eq:PPS_As_expansion}
\ln \mathcal{P}_\mathcal{R} = \ln A_s + (n_s-1) \ln \frac{k}{k_*} + \alpha_s \frac{1}{2!} \ln^2 \frac{k}{k_*}  +  \beta_s \frac{1}{3!} \ln^3 \frac{k}{k_*}  + \gamma_s \frac{1}{4!} \ln^4 \frac{k}{k_*}  \dots~,
\end{align}
from which we can derive the following relations \cite{Planck:2013jfk}:\footnote{We only give the leading order term in the potential slow-roll expansion for every coefficient.}
\begin{align}
A_{\mathrm{s}} &= \frac{V}{24\pi^2} \frac{1}{\epsilon_V} \,,\label{eq:SR_expansion_params} \\
n_{\mathrm{s}}-1 &= 2 \eta_{V}-6 \epsilon_{V}\,, \\
\alpha_s &= 16 \epsilon_{\mathrm{V}} \eta_{V}-24 \epsilon_{\mathrm{V}}^{2}-2 \xi_{V}^{2}\,, \\
\beta_s &= -192 \epsilon_{\mathrm{V}}^{3}+192 \epsilon_{\mathrm{V}}^{2} \eta_{V}-32 \epsilon_{\mathrm{V}} \eta_{\mathrm{V}}^{2}
-24 \epsilon_{\mathrm{V}} \xi_{\mathrm{V}}^{2}+2 \eta_{V} \xi_{V}^{2}+2 \omega_{V}^{3}\,, \label{eq:SR_expansion_params_last} \\
\nonumber \vdots & 
\end{align}
where all terms are evaluated at the pivot scale $k_*$ and the corresponding\footnote{This correspondence is often expressed as $a_* H(a_*) \approx k_*$ and $\varphi_* \approx \varphi(a_*)$. However, a precise calculation leads to $a_* H(a_*) = 2\pi k_*$. While this additional factor of $2\pi$ can be ignored in quasi-flat potentials, if we were considering oscillating potentials this would become significant.} pivot field $\varphi_*$
and pivot scale factor $a_*$\,. For this expansion, the slow-roll conditions imply that successive terms should be smaller than preceding terms. The higher order terms in Eq. \eqref{eq:PPS_As_expansion} are commonly referred to as the running of the scalar index and the running of the running of the scalar index.
\dnew
Summarizing, the shape of the PPS in slow-roll can be determined via its amplitude and many running indices, which map one-to-one to the slow-roll parameters. The latter are tightly linked to the inflationary potential and its derivatives, so that bounds on these parameters would translate into bounds on $V(\varphi)$. While there exists an analytically known mapping between $V(\varphi)$ and $\mathcal{P}_\mathcal{R}(k)$ in slow-roll, it is important to perform the full numerical calculation of Eq. \eqref{eq:mukhanov} in general.
\dnew
As a remark, we highlight that while the potential Taylor expansion of Sec.  \ref{ssec:potential} represents a large class of smooth potentials and does not require an explicit slow-roll approximation, it is subject to additional constraints on inflation, such as the requirement of $\epsilon_H < 1$. Instead, the slow-roll expansion described in this section is more permissive with the details of inflation and can represent any sufficiently smooth PPS, but does not translate into direct information about the potential. Additionally, since the pivot scale is fixed to that of the CMB, the slow-roll expansion is constrained to be very smooth far away from the pivot scale of the PPS. As such, the two methods are exploring different parameter regions, determined partly by the underlying assumptions. This in turn means that various combinations of the PSR parameters can correspond to potentials which are not in slow-roll during much of their evolution, meaning that we cannot simply use the slow roll approximation of \text{Eqs.~\eqref{eq:SR_expansion_params}-\eqref{eq:SR_expansion_params_last}} to directly compare the results between these two approaches. These two analyses are thus highly complimentary, and provide different information that cannot be directly compared.

\section{Inflationary imprints on the CMB} \label{sec:th_imprints}
In this work we mainly consider two ways in which the characteristics of the PPS expressed in \fulleqref{eq:dim_pps} can affect the CMB. Firstly, we focus on the CMB anisotropy power spectra and, secondly, on CMB \SD{s}. Since the theory behind the former is very well-known, we only provide a summarized review of the topic in App. \ref{app:th_CMB}. As the latter are less well-known, in the following paragraphs we briefly discuss the role of the PPS in determining the final shape of the SD signal. The notation we adopt is based on~\cite{Lucca2019Synergy}, where the reader can find the detailed definition of the different quantities mentioned below, as well as more in-depth discussions. For further considerations on the generation of SDs from the PPS, see e.g., \cite{Chluba2011Evolution, Chluba2012CMB, Chluba2012Inflaton, Pajer2012Hydrodynamical, Chluba2013CMB, Chluba2014Teasing, Cabass2016Distortions, Chluba2016Which, Abitbol2017Prospects}.
\dnewnoindent
Within this work we decompose the total sky-averaged photon spectral intensity as
\begin{align}\label{eq:tot_signal}
	I(x) = \mathcal{B}(x) + \Delta I_{\mathrm{\Lambda CDM}}(x) + \Delta I_{\mathrm{exotic}}(x) + \Delta I_T(x)  + \Delta I_{\mathrm{SZ}}(x) + \Delta I_{\mathrm{foregrounds}}(x)  ~,
\end{align}
where $x$ is the dimensionless frequency $x = h \nu/(k_B T)$ and $T$ is a chosen reference temperature which also defines the corresponding black body (BB) spectrum $\mathcal{B}(x)$. 
\dnew
In the context of this paper, we will not be considering any exotic energy injection mechanism, and, therefore, $\Delta I_{\mathrm{exotic}}(x)=0$, but rather we will be considering only the \lcdm processes of adiabatic cooling through electron scattering and the dissipation of acoustic waves. The latter will be the dominant process impacted by the precise shape of the PPS, and thus by the inflationary potential. In full generality, it is possible to define the contribution of the \lcdm model to the total SD signal as
\begin{align}
\Delta I_{\Lambda \mathrm{CDM}}(x) = \int \mathrm{d}z \,\, G_\mathrm{th}(x,z) \frac{\mathrm{d}Q(z)/\mathrm{d}z}{\rho_\gamma(z)}~,
\end{align}
where $G_\mathrm{th}$ is a pre-computed Green's function of the linear thermalization equations \cite{Chluba2013Green}, and $dQ/dz$ is the heating, which is given by
\begin{align}\label{eq:heating_lcdm}
\mathrm{d}Q/\mathrm{d}t = - H \alpha_h T_\gamma + 4 A^2 \rho_\gamma \int \mathrm{d}k k \mathcal{P}_\mathcal{R}(k) (\partial_t k_D^{-2}) \exp(-2 (k/k_D)^2)~,
\end{align}
where $A$ is the amplitude of oscillations of the photon velocity perturbation in the tight coupling approximation, $k_D$ is the diffusion damping scale of the CMB, and $\alpha_h$ and $T_\gamma$ represent respectively the heat capacity and temperature of the photon-baryon plasma (considering $z \gg 10^3$, see \cite{Lucca2019Synergy} for the corresponding definitions), and the dimensionless PPS is defined in Eq. \eqref{eq:dim_pps}. The second term of Eq. \eqref{eq:heating_lcdm} -- which represents the dissipation of acoustic waves of scalar modes -- is sensitive to the PPS on scales between $1 - 10^4$ Mpc$^{-1}$. The dissipation of tensor modes is further discussed in \cite{Chluba2014Spectral,Kite2020Bridging}, and we leave its inclusion to future work.
\dnew
As a remark, as recently pointed out in \cite{Hart:2020voa}, the SDs induced by the cosmic recombination radiation could also play an important role in breaking parameter degeneracies and thus lowering the constraints on the cosmological parameters. Since we do not include these contributions in our analysis, the constraints presented in Sec. \ref{sec:res} can only improve once the full set of SD effects predicted by the $\Lambda$CDM model is considered.
\dnew
The remaining three terms in Eq. \eqref{eq:tot_signal} represent the contributions from temperature shifts, late-time sources, and foregrounds, respectively. In order to observe the \tquote{primordial} signal ($\Delta I_{\Lambda \mathrm{CDM}}$), these need to be marginalized over.
\dnew
The first uncertainty we have to account for is given by the difference between the arbitrarily chosen reference temperature and the actual CMB temperature. Such a temperature difference can be caused by the uncertainty in the measurement of the monopole temperature today and is therefore unavoidable.\footnote{Note that energy injections in the very early universe can also change the BB temperature of the CMB, but this effect will mainly be relevant at very high redshifts, and can safely be set to zero for present-day observations \cite{Lucca2019Synergy}.} For this reason, we marginalize over possible temperature shifts $\Delta_T= \Delta T/T$ according to
\begin{align}
\Delta I_T(x) = \mathcal{N}x^3 \left(\Delta_T (1+\Delta_T) G(x) + \frac{\Delta_T^2}{2} Y(x)\right)\,,
\end{align}
with $\mathcal{N} = 2 (k_B T)^3/(hc)^2$ , $G(x) = xe^x/(e^x-1)^2$ and $Y(x)=G(x)[x(e^x+1)/(e^x-1)-4]$, as described in \cite{Lucca2019Synergy} (see in particular Sec. 3.3 of the reference for more details).
\dnew
The second contribution we focus on comes from the (thermal and kinematic) Sunyaev-Zeldovich (SZ) effect \cite{Zeldovich1969Interaction}. This effect takes place in the late universe, when reionized electrons have energies higher than the CMB photons and are able to up-scatter the latter, thus creating SDs. The main contributions to this effect arise from cosmic reionization, the intracluster medium (ICM), and the intergalactic medium (IGM). Since the averaged value of the electron temperature $T_e$ and of the optical depth $\Delta \tau$ in these environments is hard to determine, we do not treat the thermal and kinematic SZ terms fully, but instead simply marginalize over $y$ distortions from reionization, i.e., 
\begin{align}
\Delta I_\mathrm{SZ}(x) \approx \mathcal{N} x^3 \left( y_{\rm reio} Y(x) \right)\,,
\end{align}
where we neglect additional relativistic contributions (as well as any kinematic or thermal SZ contributions beyond this, see \cite{Lucca2019Synergy} for additional discussions).
In this case, we adopt a Gaussian prior on the reionization parameter $y_{\rm reio}$ with fiducial value ${y_{\rm reio}=1.77\times10^{-6}}$ as found by \cite{Hill2015Taking}, and with an uncertainty of $1.57\times10^{-6}$ as estimated in \cite{Dolag:2015dta}. This procedure has the added advantage of also marginalizing over the $y$ signal created by the CMB dipole and higher multipoles.
\dnew
The final contribution to the marginalization that we have to take into consideration comes from the sum of all foregrounds. Here we base our prescription of the foregrounds partly on the Planck results \cite{Adam2015PlanckX, Akrami2018PlanckIV} and partly on the subsequent approximations of these presented in \cite{Abitbol2017Prospects}, which have also been used in \cite{Mukherjee:2019pcq}. In the frequency range between 10 GHz and 3 THz, the main contributions are due to the presence of Galactic thermal dust and the Cosmic Infrared Background (CIB), as well as from synchrotron, free-free, spinning dust, and integrated CO emissions (see e.g. Fig. 2 of \cite{Abitbol2017Prospects} for a nice graphical overview).
\dnew
Of these foregrounds, the first two can be described by a BB spectrum multiplied by a power law and can be fully parametrized using three parameters: one amplitude $A$, one spectral index $\beta$, and one temperature $T$, taking the form
\begin{align}
	\Delta I_i(x) =  A_i \left(\frac{x_i}{x_{i,\mathrm{ref}}}\right)^{\beta_i+3}
	\frac{\exp\left(x_{i,\mathrm{ref}}\right)-1}{\exp\left(x_i\right)-1}~,
\end{align}
where $i\in \{\mathrm{thermal~dust, CIB}\}$,  $x_i=h \nu/(k_B T_i)$, and $\nu_\mathrm{ref} = 545$~GHz.\footnote{In order to describe the contribution from thermal dust, we refer to the corresponding equation given in Tab. 4 of \cite{Adam2015PlanckX}. Although a similar equation is given in Tab. 1 of \cite{Abitbol2017Prospects}, the formulation provided in \cite{Adam2015PlanckX} includes the presence of a reference frequency. This quantity is particularly important as it allows a direct physical interpretation of the spectrum amplitude, since it is simply the value of the correction at the reference frequency.} In the case of thermal dust, the spectral index and the temperature read respectively $\beta_D=1.53\pm0.05$ and $T_D=21\pm2$~K, as in Tab. 5 of \cite{Adam2015PlanckX} (see also e.g., \cite{Erler:2017dok} for related discussions). Furthermore, in the case of CIB, we have $\beta_{\rm CIB}=0.86\pm0.12$ and $T_{\rm CIB}=18.8\pm1.2$~K (see e.g., App. F of \cite{Akrami2018PlanckIV} for an interesting discussion on how this parameter compares to the Planck 2018 analysis), where the fiducials are taken from \cite{Abitbol2017Prospects} and for the uncertainties we adopt the values given in \cite{Fixsen:1998kq}. Note that while we use these measured values as priors on our spectral shapes, we do not impose any additional constraints on the amplitudes since they might be subject to large systematic calibration uncertainties for different experiments.
\dnew
Note that in the considered frequency range from $15\mathrm{GHz}$ to $1\mathrm{THz}$ both contributions are almost perfect power laws, generating a large degeneracy between the two amplitudes and the two spectral indices (see e.g., Fig. A1 of \cite{Abitbol2017Prospects} for a graphical representation). Since the variation of the thermal dust parameters almost perfectly mimics the variation of the CIB parameters, here we choose to fix the CIB parameters to their best-fit values from Table 1 of \cite{Abitbol2017Prospects}.
\dnew
Furthermore, the synchrotron emission can be described by means of a power law with logarithmic curvature involving three free parameters: one amplitude $A$, one spectral index $\alpha$, and one spectral curvature $\omega$ \cite{Abitbol2017Prospects}. In this case we have
\begin{align}
	\Delta I_\mathrm{sync}(x) = A_S \left(\frac{x_\mathrm{ref}}{x}\right)^{\alpha_S+\frac{1}{2}\omega_S \log^2({x/x_\mathrm{ref}})}\,,
\end{align}
where $x$ is defined as above and in this case $\nu_\mathrm{ref} = 100$ GHz. As suggested in \cite{Abitbol2017Prospects}, we fix the fiducial values of the amplitude and the spectral index to the value given in Tab. 1 therein with a 10\% Gaussian prior on both parameters. The remaining parameter, the curvature index, can either be set to zero (as done in \cite{Mukherjee:2019pcq}), fixed to the fiducial value of $0.2$ (as done in \cite{Abitbol2017Prospects}), or marginalized over. Within this work, we adopt the latter option.
\dnew
The next significant contribution is the SD of free-free interactions, for which we follow Tab. 4 of \cite{Adam2015PlanckX}, adopting
\begin{align}
\Delta I_\mathrm{free-free}(x) &= A_{ff} \mathcal{N} T_e (1-\exp(-\tau_{ff}))~, \\
\tau_{ff} &\approx 0.05468\, \mathrm{EM} \left(\frac{T_e}{K}\right)^{-3/2} \left(\frac{\nu}{\mathrm{GHz}}\right)^{-2} g_{ff}~,\\
g_{ff} &\approx \log\left\{e+\exp\left[5.96-\frac{\sqrt{3}}{\pi} \log\left(\left(\frac{\nu}{\mathrm{GHz}}\right) \left(\frac{T_e}{10^4K}\right)^{-3/2}\right)\right]\right\}~,
\end{align}
(see the aforementioned reference for a definition of the many quantities involved). As an additional remark, note that a similar solution has also been employed in \cite{Mukherjee:2019pcq}. However, there the emission measure $\mathrm{EM}$ has been set to 1 and instead an overall amplitude parameter has been used (for more details on the definition of $\mathrm{EM}$ see e.g., \cite{Draine:1997tb, 2011Draine}).
\dnew
The remaining contributions considered in this work are given by spinning dust and integrated CO emissions, for which we adopt spectral templates. For the case of spinning dust emission we use the same curve as in \cite{Adam2015PlanckX}, although converted from brightness temperature to intensity, while for the CO emission template we follow \cite{Abitbol2017Prospects}.
\section{The numerical implementation}\label{sec:num}
For the numerical evaluation of the inflationary quantities described in \secref{sec:th_inf}, the CMB power spectra, and the \SD signal discussed in \secref{sec:th_imprints}, we rely on the Boltzmann solver \textsc{class}~\cite{Lesgourgues2011Cosmic, Blas2011Cosmic}. In particular, we employ the version \textsc{v3.0} which includes the framework recently introduced in \cite{Lucca2019Synergy} for the calculation of CMB \SD{s} (see also \cite{Fu:2020wkq} for related discussions) as well as several options for the calculation of the inflationary potential~\cite{Lesgourgues:2007gp}.
\dnew
In order to derive the corresponding cosmological constraints on the inflationary quantities, we make use of the parameter extraction code \textsc{MontePython} \cite{Audren2013Conservative, Brinckmann2018MontePython}. This tool is particularly useful when dealing with \SD missions (and more in general with future probes)  as it allows to perform the MCMC parameter scans with so-called mock likelihoods. These likelihoods are designed to realistically emulate the observational constraints of the considered cosmological probe, but they depend on an arbitrary fiducial model which might not correspond to the actual measured results (for more details on this type of approach see e.g., \cite{Perotto2006Probing, Brinckmann2018Promising, Lucca2019Synergy}). However, by imposing the same fiducial values it is, therefore, possible to combine and compare the constraining power of different (completed or future) missions.
\dnew
Within this work we consider several such likelihoods. First of all, we will rely on completed missions such as Planck \cite{Aghanim2018PlanckVI} and FIRAS \cite{Fixsen1996Cosmic}.\footnote{Since we are using mock likelihoods for future experiments, we also used mock likelihoods even for these already completed experiments, such that the same fiducial model can be used throughout.} These are then complemented by upcoming CMB anisotropy missions such as CMB-S4 \cite{Abazajian2016CMB, Abazajian2019CMB} and LiteBIRD \cite{Matsumura2013Mission,Suzuki2018LiteBIRD}, as well as proposed \SD missions such as PIXIE \cite{Kogut2011Primordial} and PRISM \cite{Andre2014Prism}. For more details on the implementation of these CMB SD and anisotropy mock likelihoods we refer the interested reader to Section 3.3 of \cite{Lucca2019Synergy} and references \cite{Perotto2006Probing, Brinckmann2018Promising}, respectively. For all the aforementioned probes, the fiducial observables have been created in the basis $\{h,\omega_{\rm b},\omega_{\rm cdm},A_s,n_s, z_{\rm reio}\}$ with values set to the Planck 2018 best-fits \cite{Aghanim2018PlanckVI}.\footnote{Explicitly, we use $h=0.6766$, $\omega_{\rm b}=0.02242$, $\omega_{\rm cdm}=0.11933$, $A_s=2.105 \cdot 10^{-9}$, $n_s=0.9665$, $z_\mathrm{reio}=7.82$ for the  fiducial cosmology.}
\dnew
In the treatment of the \SD likelihoods considered in this work there is an important improvement with respect to the setup presented in \cite{Lucca2019Synergy, Fu:2020wkq}. In fact, in these references the contamination from galactic and extra-galactic foregrounds, as well as from late-time sources, has been neglected. However, in order to realistically evaluate an eventual \SD signal, it is vital to accurately model the various frequency dependent foregrounds that will appear for any spectroscopic instrument as well as accounting for the eventual uncertainties in the physical quantities involved in the modeling. 
\dnew
Therefore, to estimate how this would impact the constraints presented in Sec. \ref{sec:res} we can evaluate how much the $\mu$-distortion constraining power of a PIXIE-like experiment is impacted by various levels of marginalization. In fact, for a PIXIE-like experiment without any marginalization our mock likelihood has a constraining power of $\delta \mu \sim 9.4\cdot 10^{-9}$, while adding at least the temperature shift marginalization worsens it to $\delta \mu \sim 1.7\cdot 10^{-8}$, which is very close to the $\delta \mu \sim 2 \cdot 10^{-8}$ cited in \cite{Kogut2011Primordial}. Adding the simplified marginalization over reionization in terms of a $y_\mathrm{reio}$ parameter degrades these constraints down to about ${\delta \mu \sim 2.8 \cdot 10^{-8}}$. Adding all other foregrounds of Sec. \ref{sec:th_imprints} as well degrades the constraining power further to $\delta \mu \sim 8.4 \cdot 10^{-7}$. Therefore, for the present analysis, we have improved the \SD mock likelihoods to account for all of these contributions as described in Sec. \ref{sec:th_imprints}.
\dnew
Finally, we remark that, as already mentioned in Sec. \ref{sec:th_imprints}, when marginalizing over the reionization degradation we neglect the kinematic SZ effect and the relativistic contribution to the thermal SZ effect. The reason for this is that the precise values and priors of the optical depth $\Delta \tau$ and the temperature $T_e$ of the electron plasma (see Sec. 2.4.3 of \cite{Lucca2019Synergy} for more details on the impact of these quantities on the SZ signal) are not strongly constrained in the literature (see e.g. \cite{Chluba2012Fast, Chluba2013Sunyaev, Hill2015Taking, Erler:2017dok, Remazeilles2019Neglect, Chluba2019Voyage}), while the value of $y_\mathrm{reio}$ seems to be well known and modeled, as argued above. In particular, the relativistic SZ effect would need to be modeled as the superposition of many individual contributions of clusters with different $\Delta \tau, T_e, \beta, \beta_z$, where $\beta$ and $\beta_z$ respectively refer to the cluster velocity and its component along the line-of-sight. Since this would require a more detailed study, potentially including a number of hydro-dynamical simulations to obtain a suitable distribution of these parameters, we leave a more in-depth discussion for future work and refrain from using an approximate treatment in our work. However, to estimate the impact a further marginalization over these contributions would have on the measurement of the final SD signal, we added a marginalization over the optical depth and temperature of a fiducial cluster with $\beta=\beta_z=1/300$ according to the prescription of \cite{Chluba2012Fast,Chluba2013Sunyaev}, the implementation of which was already presented in \cite{Lucca2019Synergy}. The corresponding sensitivity reduces to $\delta \mu \sim 9.6 \cdot 10^{-7}$, which is only about 15\% worse than with the simplified $y_\mathrm{reio}$ treatment (an impact orders of magnitude below the galactic and extra-galactic contributions). We can then safely neglect these contributions in the reminder of this work, leaving a more advanced treatment of the reionization marginalization beyond $y_\mathrm{reio}$ for future work.
\section{Results}\label{sec:res}
In this section we apply the numerical setup introduced in Sec. \ref{sec:num} to several inflationary scenarios following the theoretical framework discussed in Sec. \ref{sec:th_inf}. In particular, in Sec.~\ref{ssec:results_potential}  we first consider a Taylor expansion of the inflationary potential up to fourth and sixth order, with particular attention to the increasing improvement obtained with the addition of SDs when including higher order terms. Then in Sec.~\ref{ssec:results_SR} we present constraints for the slow-roll expansion discussed in Sec.~\ref{ssec:th_slow_roll} up to order $\delta_s$\,. Finally, in Sec.~\ref{ssec:results_feature} we consider more exotic forms of the PPS, allowing for steps and kinks, as well as a two-component inflationary model.
\dnew
In order to evaluate the relative improvement when combining CMB with SD missions, we use two different measures. First, we quote the percentage improvement on the 1$\sigma$ (68.3\% CL) parameter constraints for each parameter separately. The corresponding formula for the improvement factor $f$ is
\begin{equation}
	f = \left[\left(\frac{\sigma_\mathrm{CMB}(\mathrm{P})}{\sigma_\mathrm{CMB+SD}(\mathrm{P})} - 1\right) \cdot 100\right] \%~,
\end{equation}
where $\sigma_x(\mathrm{P})$ is the parameter uncertainty from a mission $x$ on the parameter $\mathrm{P}$. Second, we are going to display the figure of merit (FOM) for each of the experiments, as it nicely summarizes the overall expected improvement in the constraining power of an experiment on all parameters simultaneously. The FOM \cite{Albrecht2006Report,Wang2008Figure,Amendola2012Cosmology,Majerotto2012Probing} is defined as
\begin{equation}
	\mathrm{FOM} = \frac{1}{\sqrt{\det C}}~,
\end{equation}
where $C$ is the covariance matrix of the measurement.\footnote{We exclude Planck and SD nuisance parameters from the definition of the FOM to keep the parameter basis the same. This is done such that the addition of another experiment does not artificially enhance the FOM.} Reduced error bars lead naturally to a higher FOM.

\begin{figure}
	\centering
	\includegraphics[width=9.5 cm]{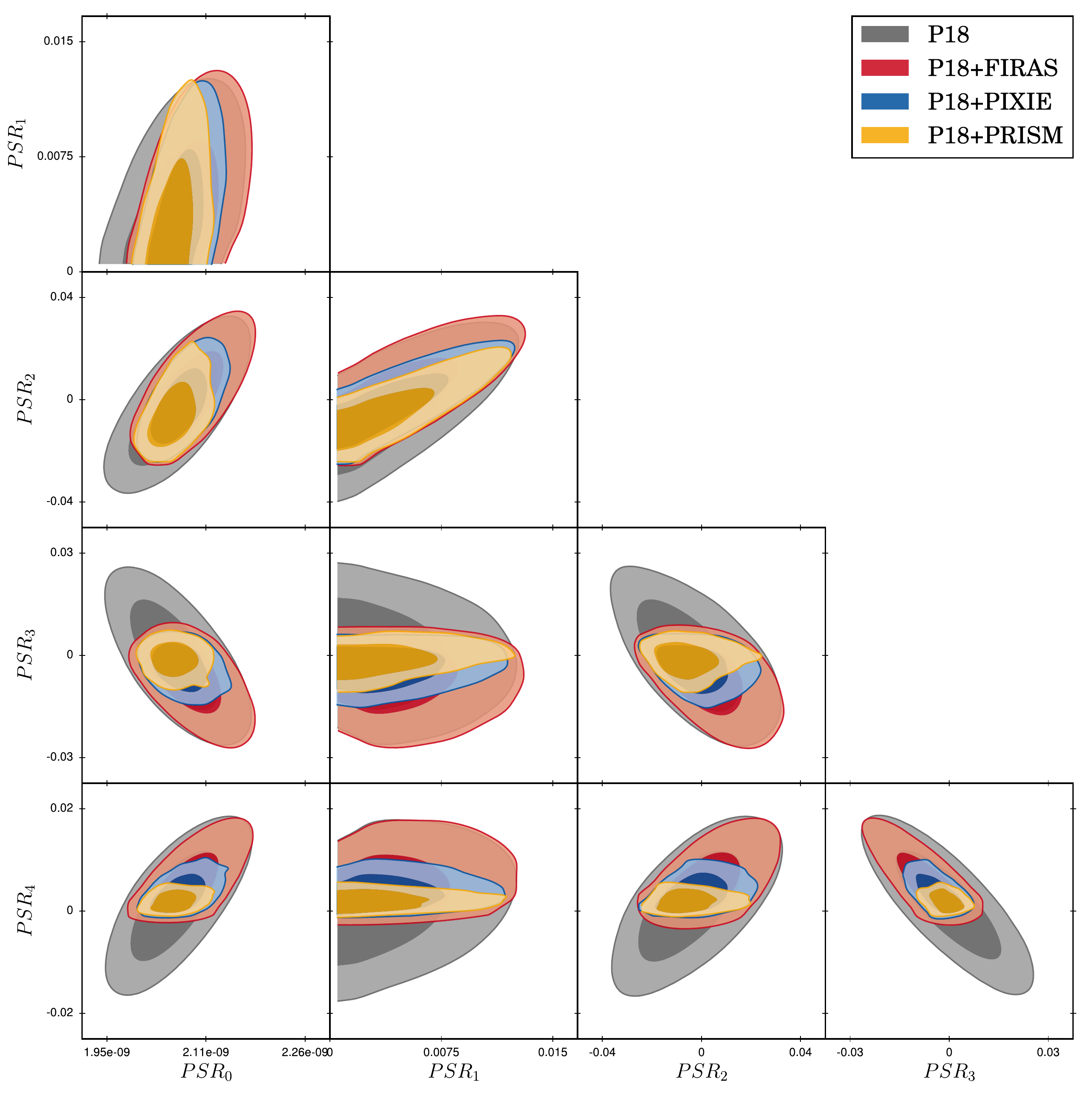}
	\caption{Posterior distributions (68.3\% and 95.4\% CL) of the PSR parameters characterizing the Taylor expansion of the inflationary potential in the V4 case for the case of Planck(+SD). In the legend we refer to Planck 2018 as P18 for brevity.}
	\label{fig:p18_V4}
	
	\vspace{1em}
	
	\begin{tabular}{l|c||c|c|c}
		 & Planck 2018 & FIRAS (\%) & PIXIE (\%) & PRISM (\%)\\ \hline
		 &&&\\[-1em]
		$\sigma(\mathrm{PSR}_0)$ & $4.3 \cdot 10^{-11}$ & 22 & 53 & 86 \\
		$\sigma(\mathrm{PSR}_1)$ & $2.6 \cdot 10^{-3}$ & -7 & 0.4 & 15\\
		$\sigma(\mathrm{PSR}_2)$ & $1.3 \cdot 10^{-2}$ & 17 & 42 & 69 \\ 
		$\sigma(\mathrm{PSR}_3)$ & $1.0 \cdot 10^{-2}$ & 52 & 110 & 210 \\
		$\sigma(\mathrm{PSR}_4)$ & $6.7 \cdot 10^{-3}$ & 70 & 170 & 350 \\ \hline \hline\rule{0pt}{2.5ex}
		FOM & $6.3 \cdot 10^{31}$ & 50 & 190 & 540 \\ \hline
	\end{tabular}
	\captionof{table}{\textbf{Top panel}: 1$\sigma$ (68.3\% CL) uncertainties on the PSR parameters for the V4 case assuming Planck(+SDs). The second column denotes the uncertainties obtainable with Planck, while the three columns on the right list the improvement/degradation given by the addition of the respective SD mission. \textbf{Bottom panel}: As in top panel but with respect to the FOM.
	\label{tab:improvements_p18_V_4}}
\end{figure}

\begin{figure}
	\centering
	\includegraphics[width=9.5 cm]{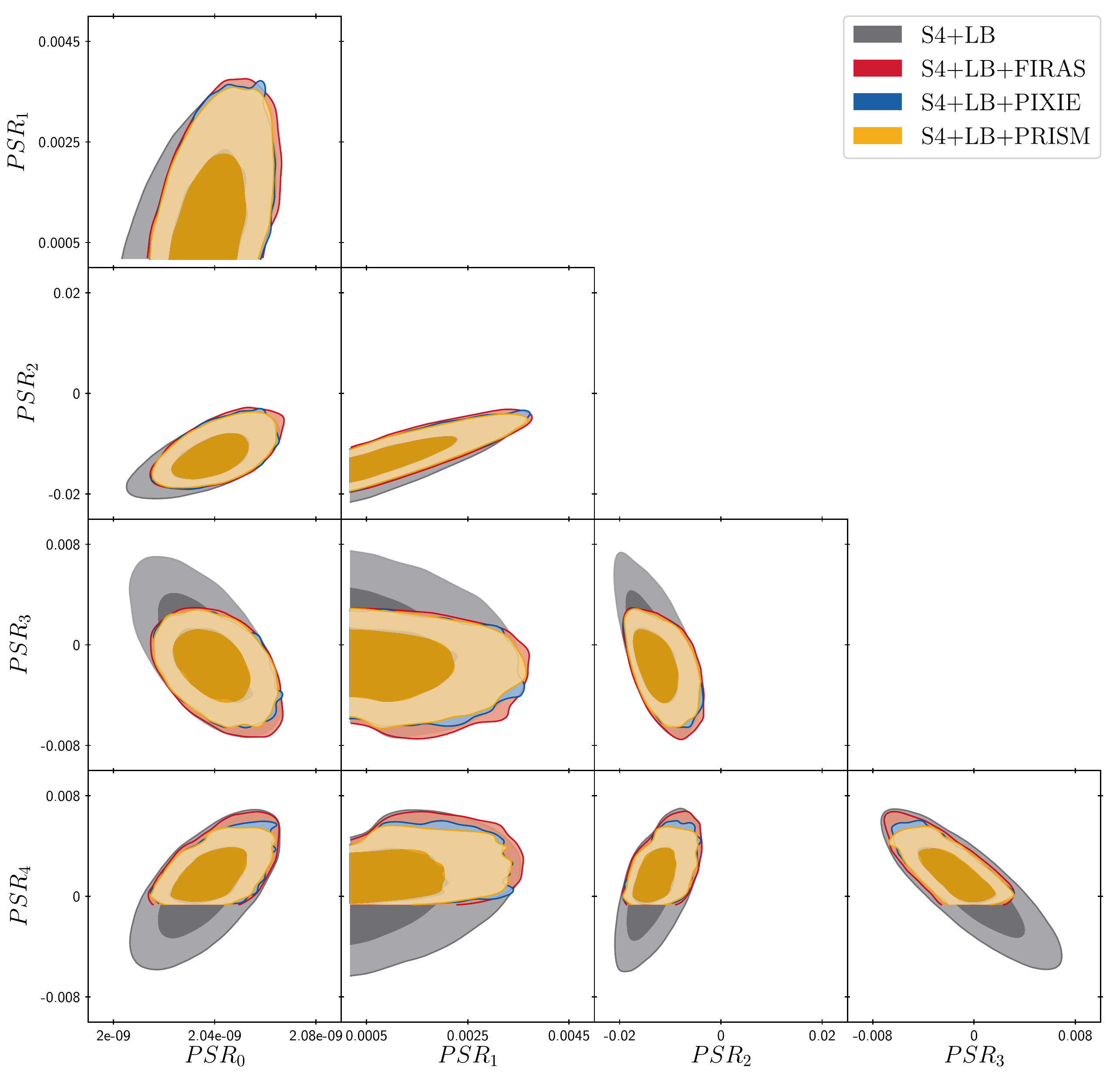}
	\caption{Same as in Fig. \ref{fig:p18_V4}, but for the combination of CMB-S4+LiteBIRD(+SD). In the legend we refer to CMB-S4 and LiteBIRD respectively as S4 and LB for brevity.}
	\label{fig:s4lb_V4}
	
	\vspace{1em}
	\begin{tabular}{l|c ||c|c|c}
		 & CMB-S4+LiteBIRD & FIRAS (\%) & PIXIE (\%) & PRISM (\%)\\ \hline
		 &&&\\[-1em]
		$\sigma(\mathrm{PSR}_0)$ & $1.2 \cdot 10^{-11}$ & 13 & 15 & 18 \\
		$\sigma(\mathrm{PSR}_1)$ & $7.1 \cdot 10^{-4}$ & -10 & -8 & -7 \\
		$\sigma(\mathrm{PSR}_2)$ & $3.4 \cdot 10^{-3}$ & 5 & 9 & 12 \\ 
		$\sigma(\mathrm{PSR}_3)$ & $2.9 \cdot 10^{-3}$ & 40 & 46 & 50 \\
		$\sigma(\mathrm{PSR}_4)$ & $2.5 \cdot 10^{-3}$ & 60 & 69 & 71 \\ \hline \hline\rule{0pt}{2.5ex}
		FOM & $2.4\cdot 10^{36}$ & 36 & 59 & 62 \\ \hline
	\end{tabular}
	\captionof{table}{Same as in Tab. \ref{tab:improvements_p18_V_4}, but for the combination of CMB-S4+LiteBIRD(+SD). \label{tab:improvements_s4lb_V_4}}
\end{figure}

\begin{figure}
	\centering 
	\includegraphics[width=15 cm]{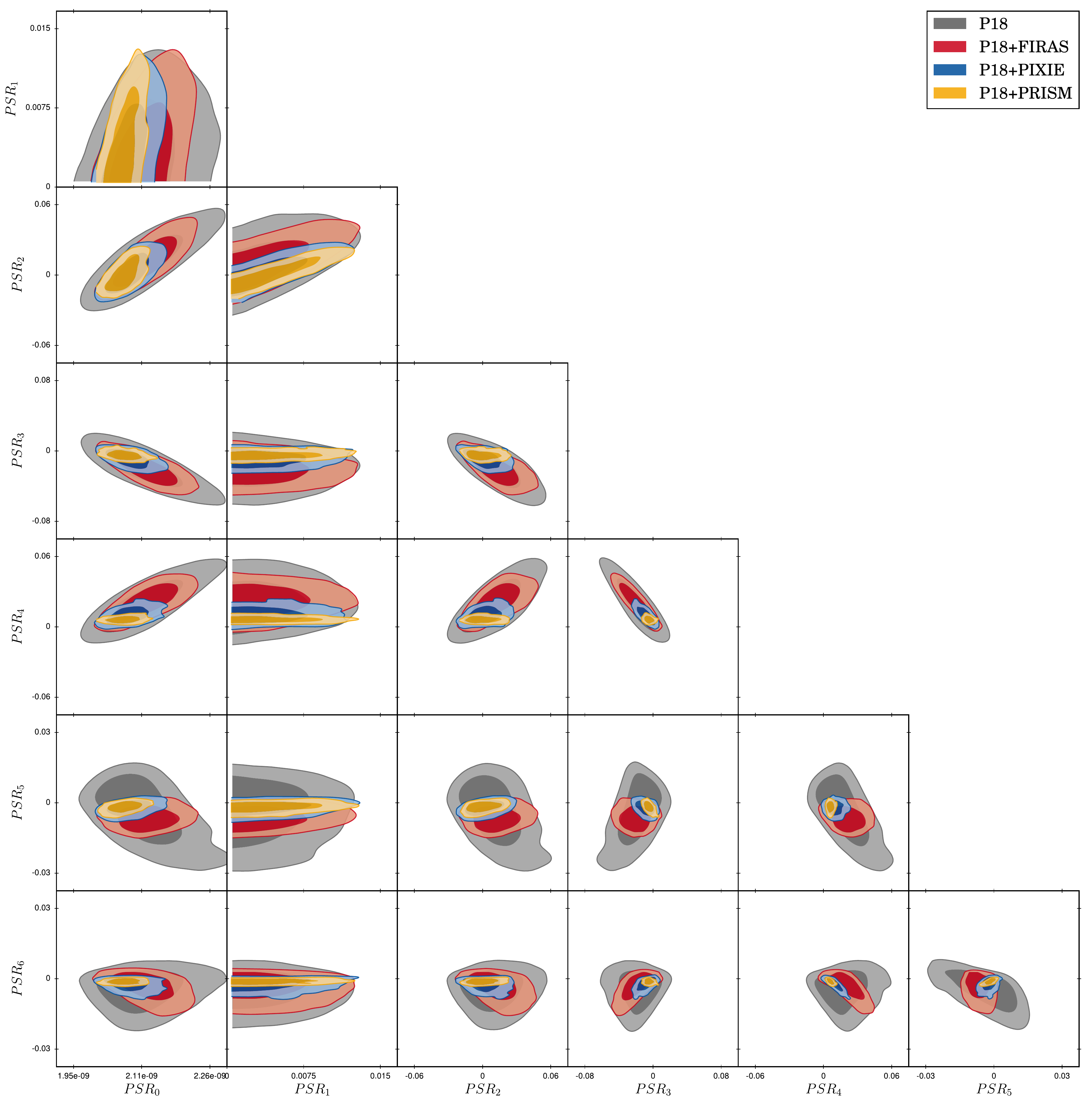}
	\caption{Same as in Fig. \ref{fig:p18_V4}, but for the V6 case.}
	\label{fig:p18_V6}
	
	\vspace{1em}
	\centering
	\begin{tabular}{l|c||c|c|c}
		 & Planck 2018 & FIRAS (\%) & PIXIE (\%) & PRISM (\%)\\ \hline
		 &&&\\[-1em]
		$\sigma(\mathrm{PSR}_0)$ & $6.1 \cdot 10^{-11}$ & 31 & 90 & 150 \\
		$\sigma(\mathrm{PSR}_1)$ & $2.6 \cdot 10^{-3}$ & 0.06 & 1.3 & -11\\
		$\sigma(\mathrm{PSR}_2)$ & $1.7 \cdot 10^{-2}$ & 24 & 59 & 79 \\ 
		$\sigma(\mathrm{PSR}_3)$ & $1.6 \cdot 10^{-2}$ & 27 & 120 & 340 \\
		$\sigma(\mathrm{PSR}_4)$ & $1.5 \cdot 10^{-2}$ & 45 & 180 & 640 \\
		$\sigma(\mathrm{PSR}_5)$ & $1.0 \cdot 10^{-2}$ & 170 & 360 & 550 \\
		$\sigma(\mathrm{PSR}_6)$ & $5.6 \cdot 10^{-3}$ & 52 & 200 & 520\\ \hline \hline\rule{0pt}{2.5ex}
		FOM & $8.1\cdot 10^{35}$ & 510 & 9400 & $1.65\cdot10^5$ \\ \hline
	\end{tabular}
	\captionof{table}{Same as in Tab. \ref{tab:improvements_p18_V_4}, but for the V6 case. \label{tab:improvements_p18_V_6}}
\end{figure}

\begin{figure}
	\centering 
	\includegraphics[width=15 cm]{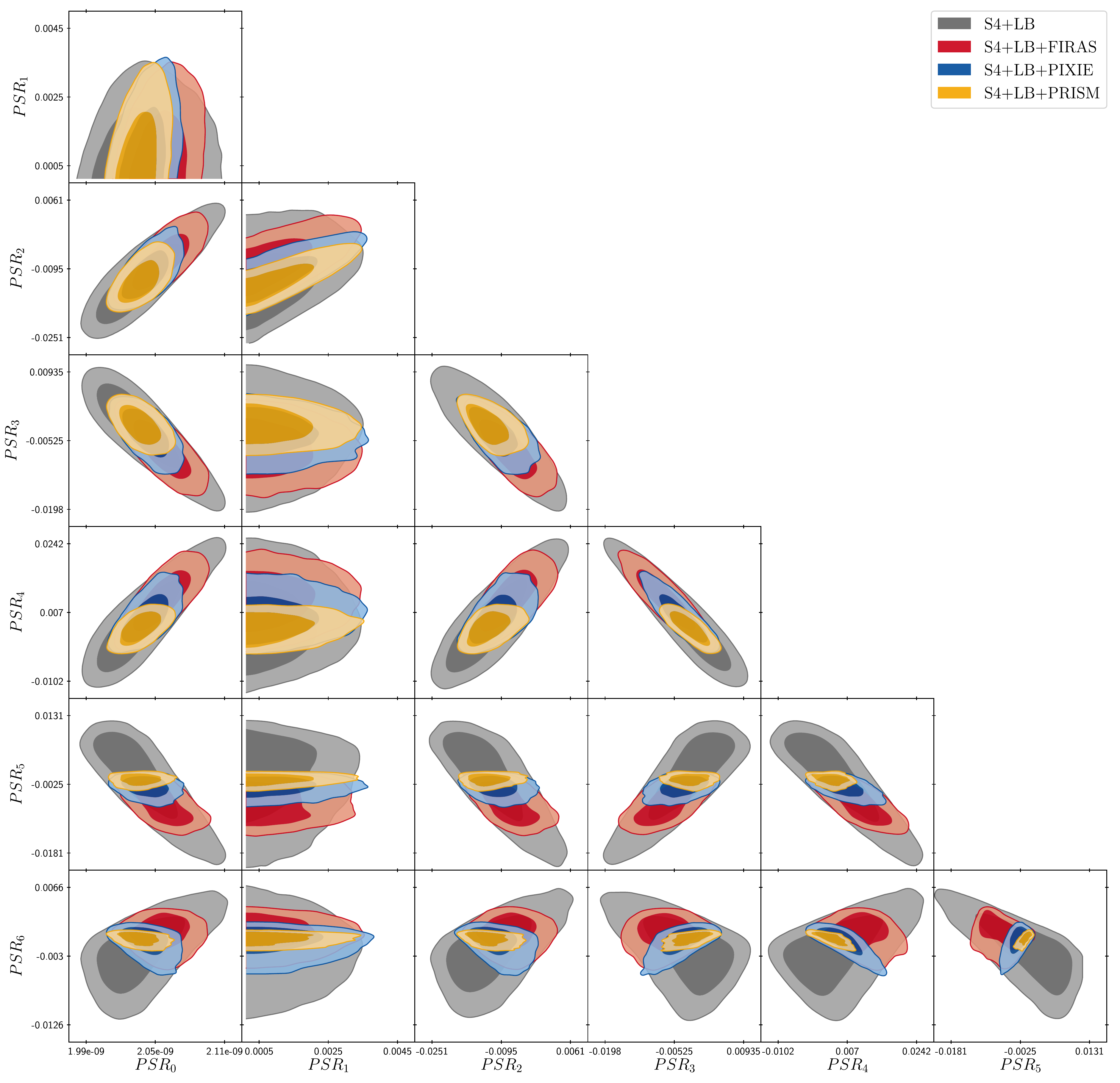}
	\caption{Same as in Fig. \ref{fig:s4lb_V4}, but for the V6 case.}
	\label{fig:s4lb_V6}
	
	\vspace{1em}
	\centering
	\begin{tabular}{l|c||c|c|c}
		& CMB-S4+LiteBIRD & FIRAS (\%) & PIXIE (\%) & PRISM (\%)\\ \hline
		&&&\\[-1em]
		$\sigma(\mathrm{PSR}_0)$ & $2.3 \cdot 10^{-11}$ & 43 & 78 & 101 \\
		$\sigma(\mathrm{PSR}_1)$ & $6.8 \cdot 10^{-4}$ & -2.2 & -4.2 & 0.27 \\
		$\sigma(\mathrm{PSR}_2)$ & $6.0 \cdot 10^{-3}$ & 42 & 71 & 94 \\ 
		$\sigma(\mathrm{PSR}_3)$ & $5.8 \cdot 10^{-3}$ & 39 & 73 & 112 \\
		$\sigma(\mathrm{PSR}_4)$ & $7.5 \cdot 10^{-3}$ & 48 & 93 & 173 \\
		$\sigma(\mathrm{PSR}_5)$ & $6.5 \cdot 10^{-3}$ & 88 & 314 & 627 \\
		$\sigma(\mathrm{PSR}_6)$ & $3.4 \cdot 10^{-3}$ & 111 & 187 & 454\\ \hline \hline\rule{0pt}{2.5ex}
		FOM & $9.7\cdot 10^{40}$ & 35 & 1409 & $1.2\cdot10^4$ \\ \hline
	\end{tabular}
	\captionof{table}{Same as in Tab. \ref{tab:improvements_s4lb_V_4}, but for the V6 case. \label{tab:improvements_s4lb_V_6}}
	\vspace{0.3 cm}
\end{figure}

\subsection{Inflationary potential expansion}\label{ssec:results_potential}
The first example of inflationary scenario we consider is given by the Taylor expansion of the inflationary potential around $\varphi=0$ as introduced in Sec. \ref{ssec:potential}. As a choice, we present the results using the PSR parameters defined in Eq. \eqref{eq:SR_params} adopting the convention  $\mathrm{PSR}_1=\epsilon_{\mathrm{V}}$, $\mathrm{PSR}_2=\eta_{\mathrm{V}}$, $\mathrm{PSR}_3=\xi_{\mathrm{V}}$, $\mathrm{PSR}_4=\omega_{\mathrm{V}}$, and defining further terms as $\mathrm{PSR}_n=s^{(n)}_\mathrm{V}$. Additionally, we have set $\mathrm{PSR}_0=A_s$. Note that this modeling of the inflationary potential does not imply any assumption related to slow-roll, and we instead numerically integrate the background equation~\eqref{eq:kleingordon} and the Mukhanov equation~\eqref{eq:mukhanov}.
\dnew
Specifically, here we present the results for two possible expansions. The first is taken until fourth order ($N=4$ in Eq. \eqref{eq:Vexpansion}) in the Taylor series and will be labeled as V4 henceforth. The second is further extended to sixth order and referred to henceforth as V6. For both cases we exploit the combination of the concluded and upcoming CMB anisotropy missions Planck and CMB-S4+LiteBIRD (which represents a realistic measurement achievable in the next decade) together with three different SD missions: FIRAS, PIXIE, and PRISM.
\dnew
The results of the V4 case are shown in Fig. \ref{fig:p18_V4} for Planck and its combination with various SD missions, and in Fig. \ref{fig:s4lb_V4} for the future CMB-S4+LiteBIRD missions and combinations with SD experiments. A quantitative description of the improvements achieved by combining anistropy and SD missions is given in Tabs. \ref{tab:improvements_p18_V_4} and \ref{tab:improvements_s4lb_V_4}.
\dnew
Although SD missions with PRISM-like sensitivities have already been shown to provide tighter constraints on inflationary parameters \cite{Fu:2020wkq}, in this context at least part of the improvement is simply due to the observation of previously inaccessible scales. In fact, as explained in Sec. \ref{sec:th_imprints} and App. \ref{app:th_CMB}, CMB anisotropy measurements are sensitive to Fourier modes in the range between $10^{-4}$ Mpc$^{-1}$ and $0.5$ Mpc$^{-1}$, while SDs cover the interval of $1-10^4$ Mpc$^{-1}$. Therefore, the addition of an SD mission intrinsically allows to extend the observed range by four orders of magnitude in $k$-space. Forcing inflation to last a sufficient number of $e$-folds such that the largest wavenumber is allowed to cross the Hubble horizon sharply limits the freedom of higher order terms.\footnote{This is a consequence of the choice made in Sec.~\ref{ssec:potential} to impose no assumptions on the end of inflation.} In particular, the reason that the negative values of PSR${}_4$ are disallowed is that the concave potential for large negative PSR${}_4$ leads to growing $V'/V$ during inflation and thus quickly  saturating $\epsilon_V \sim \epsilon_H > 1$, which ends inflation. This does not happen for convex potentials for which $V'/V$ is decreasing. Thus, the concave potentials commonly lead to inflation which does not last the required number of $e$-folds for modes with large $k_\mathrm{max} = aH$. Since these large modes are mostly probed by SDs, we see the sharp difference between constraints with and without SDs.
\dnew
Nevertheless, besides this effect based purely on the observation of new scales, the SD experiments also provide an additional anchor at small scales, thereby strengthening the lever arm proportionally to the sensitivity of the missions. In Fig. \ref{fig:p18_V4} it is evident that the improved sensitivity of upcoming SD missions greatly tightens the constraints on the PSR parameters. When including any of the considered SD missions in addition to the Planck alone case, there is a strong improvement on the bounds of all of the PSR parameters by up to around 350\% for the $\mathrm{PSR}_4$ parameter.\footnote{Note that a deterioration of the constraint such as the $-7\%$ in the case of the $\mathrm{PSR}_1$ parameter for Planck+FIRAS likely occurred due to the shift in the degeneracy direction of the parameter, and should not be interpreted as the combined experiment actually weakening the overall constraint.} The FOM even increases by up to 540\% when combining Planck with PRISM.
\dnew
However, this extended lever-arm is not as relevant for the upcoming CMB-S4+LiteBIRD missions, for which the constraints within the scales observable by the CMB anisotropy experiments are already strong enough to set tight bounds for a low order expansion such as V4. In this case, the improvement coming from the extension of the lever-arm is marginal, as evident from Tab. \ref{tab:improvements_s4lb_V_4}, where the improvements on the parameter constraints of more advanced SD missions are only marginally better than those of FIRAS. However, the latter does already clearly improve on the CMB-S4+LiteBIRD case alone by up to 60\% simply because of the additional scales it explores, which is \textit{per se} still very remarkable.
\dnew
Note that the analysis without SDs for this case was performed in Ref. \cite{Akrami2018PlanckX} utilizing real Planck data. A comparison of the green dashed contours of Fig. 10 of the reference reveals a striking agreement with the gray contours present in Fig. \ref{fig:p18_V4} of this work, directly validating the mock likelihood approach.
\dnew
For higher order expansions, the improved lever-arm from SD missions becomes much more relevant. Therefore, in order to truly appreciate the improvement SDs provide on CMB anisotropy bounds, one needs to extend the potential expansion up to higher orders. 
This is due to higher order coefficients profiting more strongly from the interplay between missions probing complementary scales\footnote{We show a simplified argument for this statement in section \ref{ssec:results_SR}, noting that it applies to the potential Taylor expansion only indirectly, since the mapping between the potential expansion and the slow-roll expansion is not one-to-one.}, since even potentials that happen to have a similar shape on CMB scales can have larger deviations on smaller scales when higher order coefficients are included. With this in mind, we further extend the inflationary potential expansion to sixth order (V6). The results for this case are shown in Figs. \ref{fig:p18_V6} and \ref{fig:s4lb_V6} for the Planck(+SDs) and CMB-S4+LiteBIRD(+SDs) combinations, respectively. As in the V4 case, we also report the improvements on the uncertainties due to the addition of the SD missions in Tabs.~\ref{tab:improvements_p18_V_6} and~\ref{tab:improvements_s4lb_V_6}.
\dnew
As expected, in the V6 case the improvements brought by SDs to the higher orders are substantial. For Planck\footnote{Since we are using a mock likelihood, we note that these are realistic estimates of the errors you would get with the real Planck data.} the improvement from PRISM is up to 640\% for $\mathrm{PSR}_4$, while for PIXIE and FIRAS the maximum improvements are as strong as 360\% and 170\%, respectively. The corresponding FOM for PRISM improves even by a factor of around 1650. Furthermore, contrary to the V4 case, in this context going from Planck to CMB-S4+LiteBIRD only minimally affects the aforementioned conclusions, with improvements on the FOM by up to a factor 120. Furthermore, while in principle SDs could be expected to impose strictly positive (negative) values for even (odd) PSR parameters in the V6 scenario (as seen in the V4 case), here the inflaton field never reaches the asymptotic regime, so that there is a strong interplay between PSR${}_5$ and PSR${}_6$, and we do not see the same clear preference.
\dnew
These results clearly show that the interplay between future CMD SD and anisotropy missions would allow, in principle, to test the flatness of the inflationary potential and, consequently, of the PPS to an unprecedented degree of precision and independently of any slow-roll assumptions.

\FloatBarrier

\begin{figure}
	\centering 
	\includegraphics[width=\textwidth]{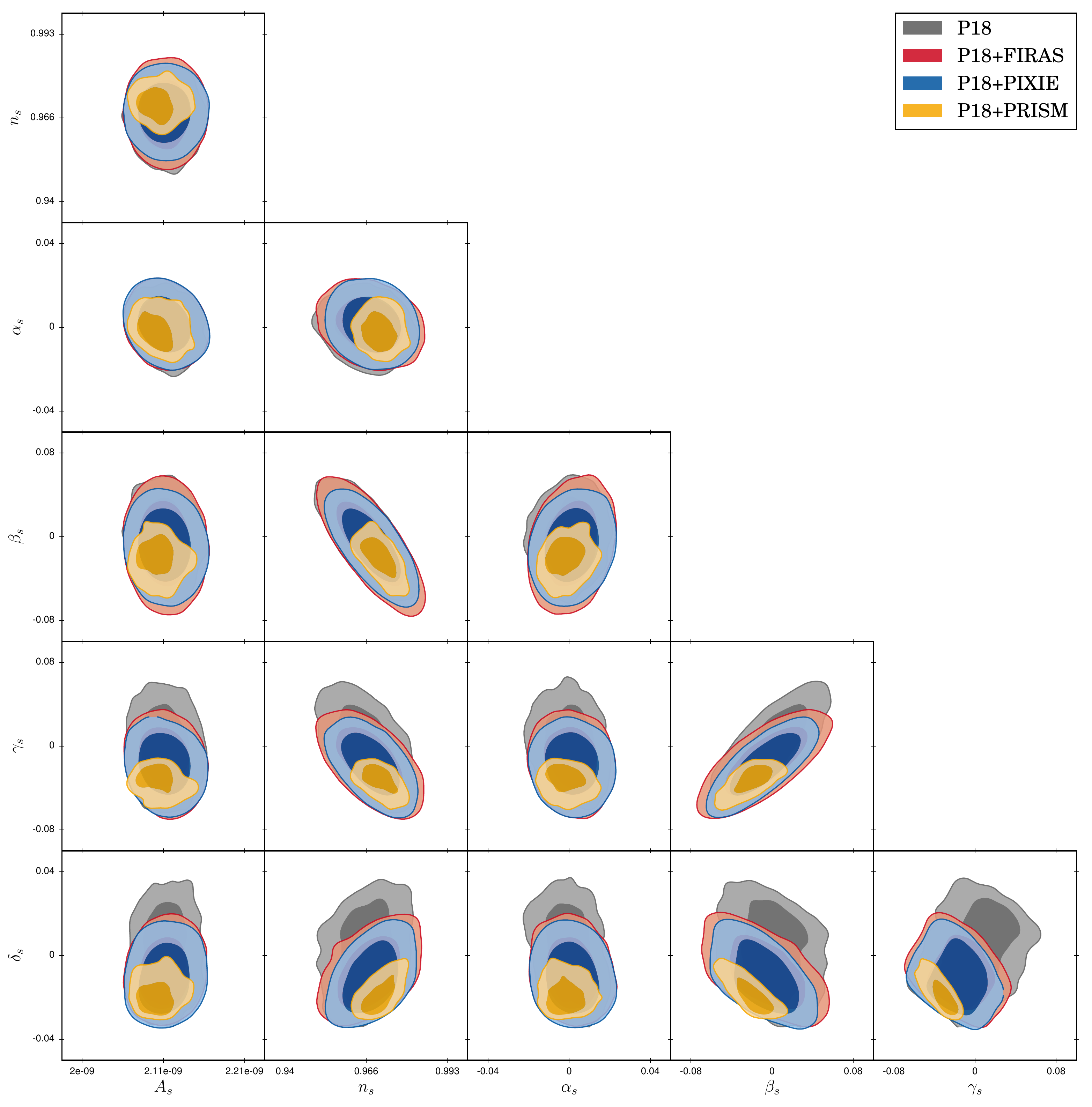}
	\caption{Same as in Fig. \ref{fig:p18_V4}, but for the slow-roll expansion case.}
	\label{fig:p18_SR}
	
	\vspace{1em}
	\begin{tabular}{l|c||c|c|c}
		& Planck 2018 & FIRAS(\%) & PIXIE (\%) & PRISM (\%)\\ \hline
		$\sigma(A_s)$ & $2.1 \cdot 10^{-11}$ & -2.0& -2.9 & 32 \\
		$\sigma(n_s)$ & $6.4 \cdot 10^{-3}$ & -7.9&4.8 & 76\\
		$\sigma(\alpha_s)$ & $8.2 \cdot 10^{-3}$ &-3.5& -2.8 & 31 \\ 
		$\sigma(\beta_s)$ & $2.4 \cdot 10^{-2}$  &-9.4& 6.7 & 96 \\
		$\sigma(\gamma_s)$ & $1.5 \cdot 10^{-2}$ &22& 34 & 200 \\
		$\sigma(\delta_s)$ & $1.6 \cdot 10^{-2}$ &47 & 47 & 200 \\ \hline \hline\rule{0pt}{2.5ex}
		FOM & $1.0 \cdot 10^{32}$ & 36 & 76 & $1.05\cdot 10^4$ \\ \hline
	\end{tabular}
	\captionof{table}{Same as in Tab. \ref{tab:improvements_p18_V_4}, but for the slow-roll expansion case.
	\label{tab:improvements_As_expansion_P18}}
\end{figure}

\begin{figure}
	\centering 
	\includegraphics[width=\textwidth]{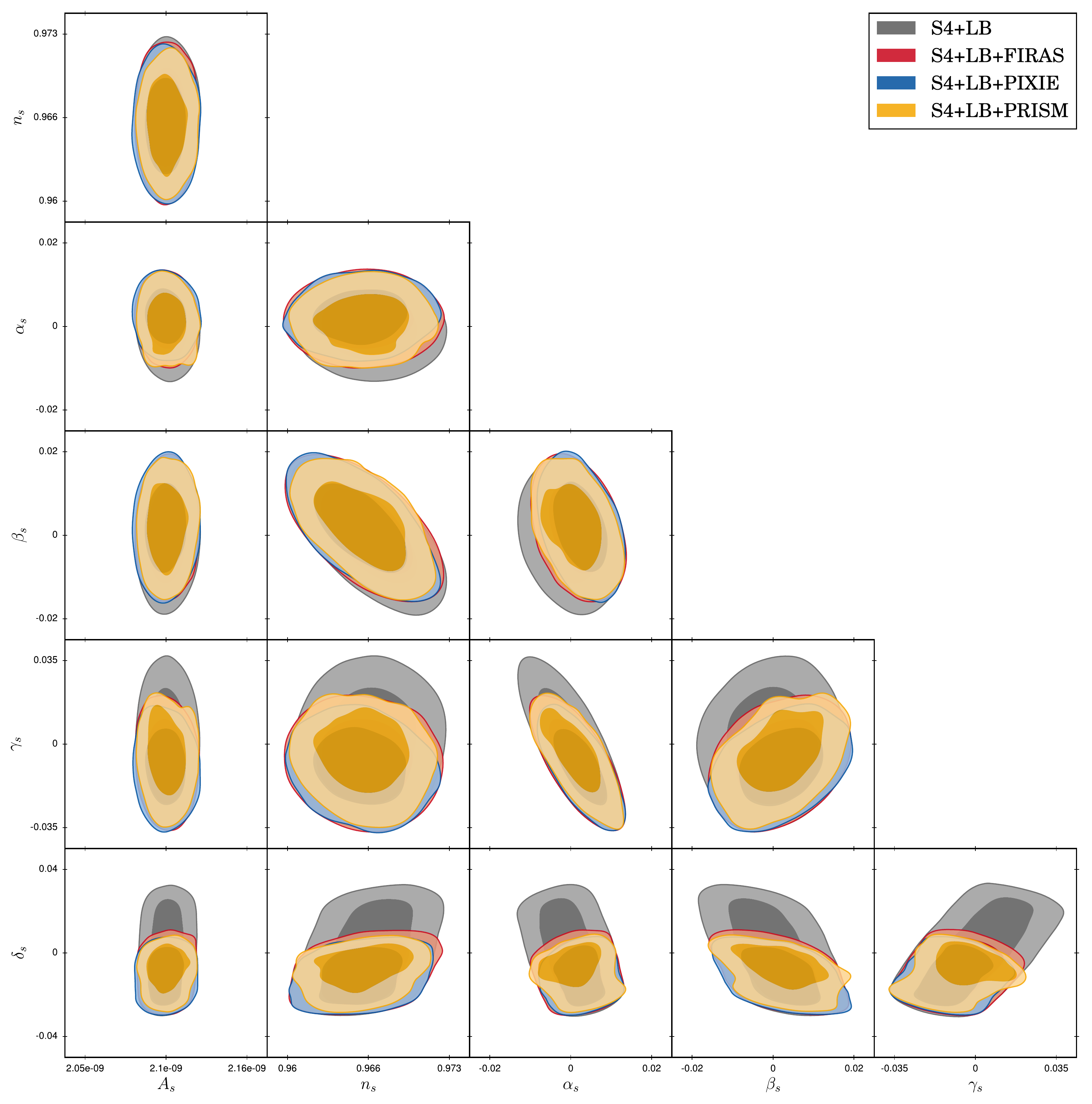}
	\caption{Same as in Fig. \ref{fig:s4lb_V4}, but for the slow-roll expansion case.}
	\label{fig:s4lb_SR}
	
	\vspace{1em}
	\begin{tabular}{l|c||c|c|c}
		& CMB-S4+LiteBIRD & FIRAS(\%) & PIXIE (\%) & PRISM (\%)\\ \hline
		$\sigma(A_s)$ & $8.6 \cdot 10^{-12}$ &  0.7 & -2.2 & -0.2\\
		$\sigma(n_s)$ & $2.5 \cdot 10^{-3}$ &  0.5 &2.4 & 5.2\\
		$\sigma(\alpha_s)$ & $5.2 \cdot 10^{-3}$ &14& 23 & 8.8 \\ 
		$\sigma(\beta_s)$ & $7.3 \cdot 10^{-3}$  &5.1& 7.4 &9.1 \\
		$\sigma(\gamma_s)$ & $1.5 \cdot 10^{-2}$ &33& 44 & 36 \\
		$\sigma(\delta_s)$ & $1.6 \cdot 10^{-2}$ &63 & 79 & 129 \\ \hline \hline\rule{0pt}{2.5ex}
		FOM & $1.6 \cdot 10^{36}$ & 69 & 77 & 189 \\ \hline
	\end{tabular}
	\captionof{table}{Same as in Tab. \ref{tab:improvements_s4lb_V_4}, but for the slow-roll expansion case
	\label{tab:improvements_As_expansion_s4lb}}
\end{figure}

\subsection{Slow-roll expansion up to $\delta_s$}\label{ssec:results_SR}
Next, we explicitly assume a slow-roll expansion to be valid, and explore its phenomenological consequences on the PPS using the expansion introduced in Eq.~\eqref{eq:PPS_As_expansion}. We run the same combinations of Planck and CMB-S4+LiteBIRD with SD experiments as in Sec. \ref{ssec:results_potential}. The results for Planck(+SDs) and CMB-S4+LiteBIRD(+SDs) are shown in Figs.~\ref{fig:p18_SR} and \ref{fig:s4lb_SR}, respectively, and summarized quantitatively in Tabs.~\ref{tab:improvements_As_expansion_P18} and \ref{tab:improvements_As_expansion_s4lb}. In the Planck case, the constraints on the leading order coefficients are barely improved, while the constraints on the higher order terms are improved by up to $200\%$. The combinations involving \text{CMB-S4+LiteBIRD} show a similar behavior, although with a strongly mitigated impact, especially for the lower order expansion parameters.
\dnew
The interpretation of this result is very similar to the discussion above: the higher order coefficients more greatly benefit from the extension of the lever-arm. This is because the constraint on the $n$-th coefficient is expected to improve approximately as the separation of the minimum and maximum scales (referred to henceforth as $\Delta$) to the power of $n$. In order to show this in a simplified manner, let us assume that each probe (CMB anisotropies or SDs) is represented by a \textit{single} data point. Then the anisotropy measurements would measure at some scale $\ln k_1 \approx \ln k_*$ some mean $\mu_1\equiv\ln \mathcal{P}_\mathcal{R}(\ln k_*) = \ln A_s$ with some uncertainty $\sigma_1=\sigma_{\ln A_s}$\,. Similarly, we would represent the SD measurement as a single data point at some smaller scale $\ln k_2$ with some mean $\mu_2 \equiv \ln \mathcal{P}_\mathcal{R}(\ln k_2)$ and corresponding uncertainty $\sigma_2$. We can thus express the scale $\ln k_2 = \ln k_1 + \Delta \approx \ln k_* + \Delta$, where $\Delta = \ln k_2 - \ln k_1$ is the separation of the minimum and maximum scales. Let us then focus on the $n$-th coefficient of the running expansion, which behaves as a function of $\ln k$ as $\ln \mathcal{P}_\mathcal{R}(\ln k)=a+c_n (\ln k/k_*)^n$ (see Eq.~\eqref{eq:PPS_As_expansion}). From a least squares regression we would find $c_n=(\mu_1-\mu_2)/\Delta^n$ while $a=\mu_1$. The uncertainty on the $n$-th coefficient of the running expansion, $c_n$, is then $\sigma_{c_n}=\sqrt{\sigma_1^2+\sigma_2^2}/\Delta^n$, which indeed scales as $\Delta^{-n}$ as anticipated above. 
\dnew
Of course, the full analysis is not quite as trivial, since each measurement actually covers a range of scales with varying sensitivity, the lower order coefficients are degenerate with other $\Lambda$CDM parameters (e.g., as in the case of the $A_s \leftrightarrow \tau_\mathrm{reio}$ degeneracy), and all coefficients have varying correlations. However, although all of these effects naturally lead to deviations from the aforementioned simple scaling, this approximate scaling is still very useful as an intuitive order of magnitude estimate of the improvements brought by the extension of the observed Fourier modes.
\dnew
These results can be compared to SD-independent constraints, such as those obtained from Planck in combination with Planck lensing \cite{Akrami2018PlanckX} or in combination with Weak Lensing and BAO data \cite{Cabass2016Constraints}. The results of both studies agree well with our Planck-only constraints up to around half an order of magnitude. Moreover, in Tabs. 2 and 5 of \cite{Akrami2018PlanckX} we can see that the constraints for lower order coefficients are only weakly impacted by including higher order coefficients, allowing for such a comparison even as the order of the expansion differs. Furthermore, in Tab. 1 of \cite{Cabass2016Constraints} it can be seen that the constraints for higher order parameters are not significantly impacted by Weak Lensing or BAO data, once again highlighting the crucial role that SDs will have in confirming and extending CMB bounds on inflation.
\dnew
Finally, note that for many inflationary models the shapes of the PPS can be mapped into an effective expansion such as in Eq. \eqref{eq:PPS_As_expansion}, and thus the forecasted constraints derived here can be used to derive corresponding constraints for the respective inflationary models. 

\subsection{Primordial power spectrum with features}\label{ssec:results_feature}

\begin{figure}
	\centering 
	\includegraphics[width=7.5cm]{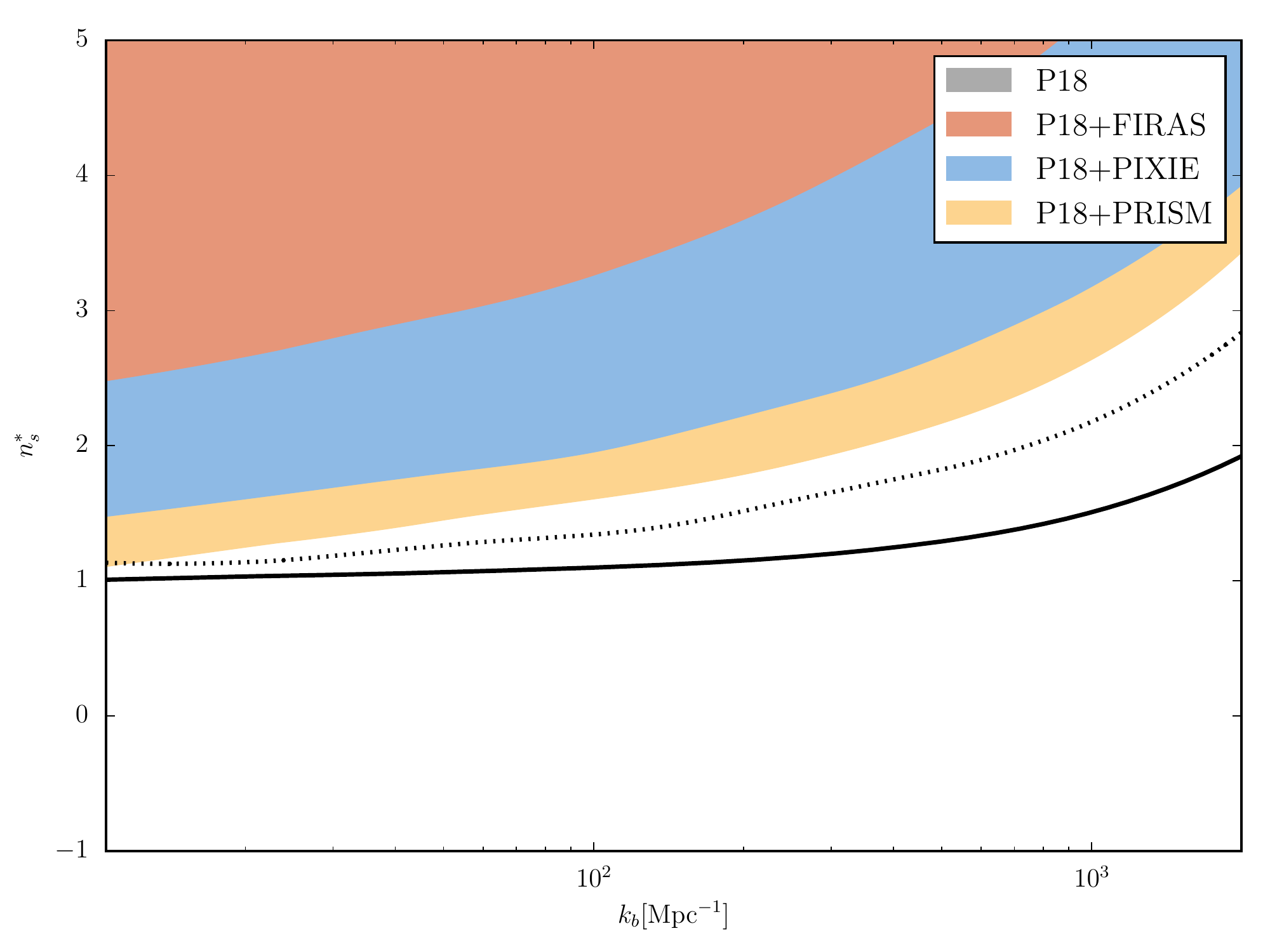}
	\includegraphics[width=7.5 cm]{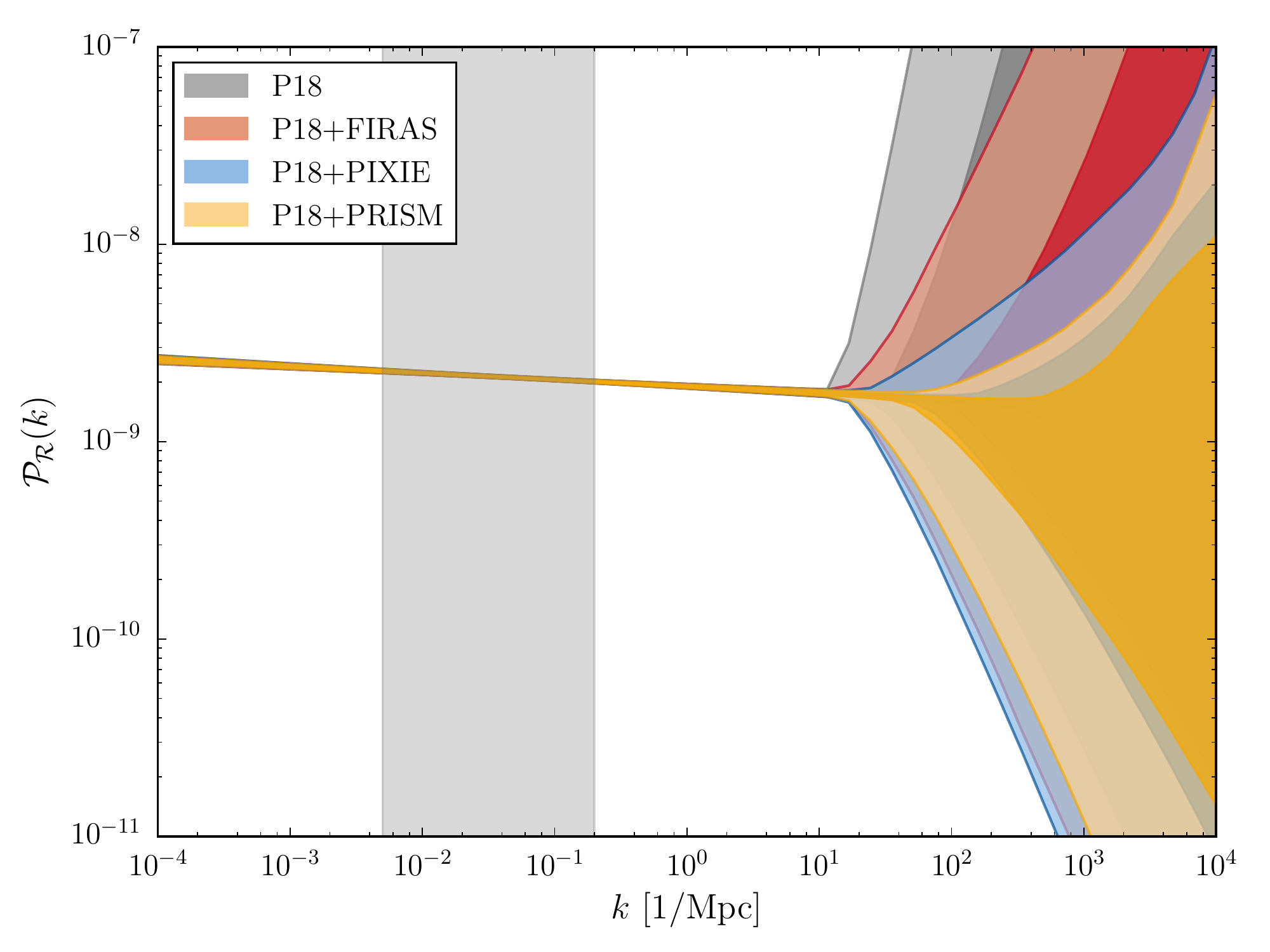}
	\caption{\textbf{Left}: Exclusion regions (95.4\% CL) for the scenario allowing for a kink in the PPS. The dotted and solid black lines represent the results for a PIXIE-like experiment only marginalizing over $\Delta_T$\, (dotted line) and assuming a simplified evaluation procedure more closely following \cite{Chluba2012Inflaton} (solid line, see text for further details). \textbf{Right}: 1$\sigma$ (68.3\% CL) and 2$\sigma$ (95.4\% CL) reconstruction of the PPS within the model. The gray band represents the scales between $5 \cdot 10^{-3}$ Mpc$^{-1}$ and 0.2 Mpc$^{-1}$, which approximately correspond to the sensitivity range of Planck. }
	\label{fig:p18_chluba}
\end{figure}

\begin{figure}[t]
	\centering 
	\includegraphics[width=7.5 cm]{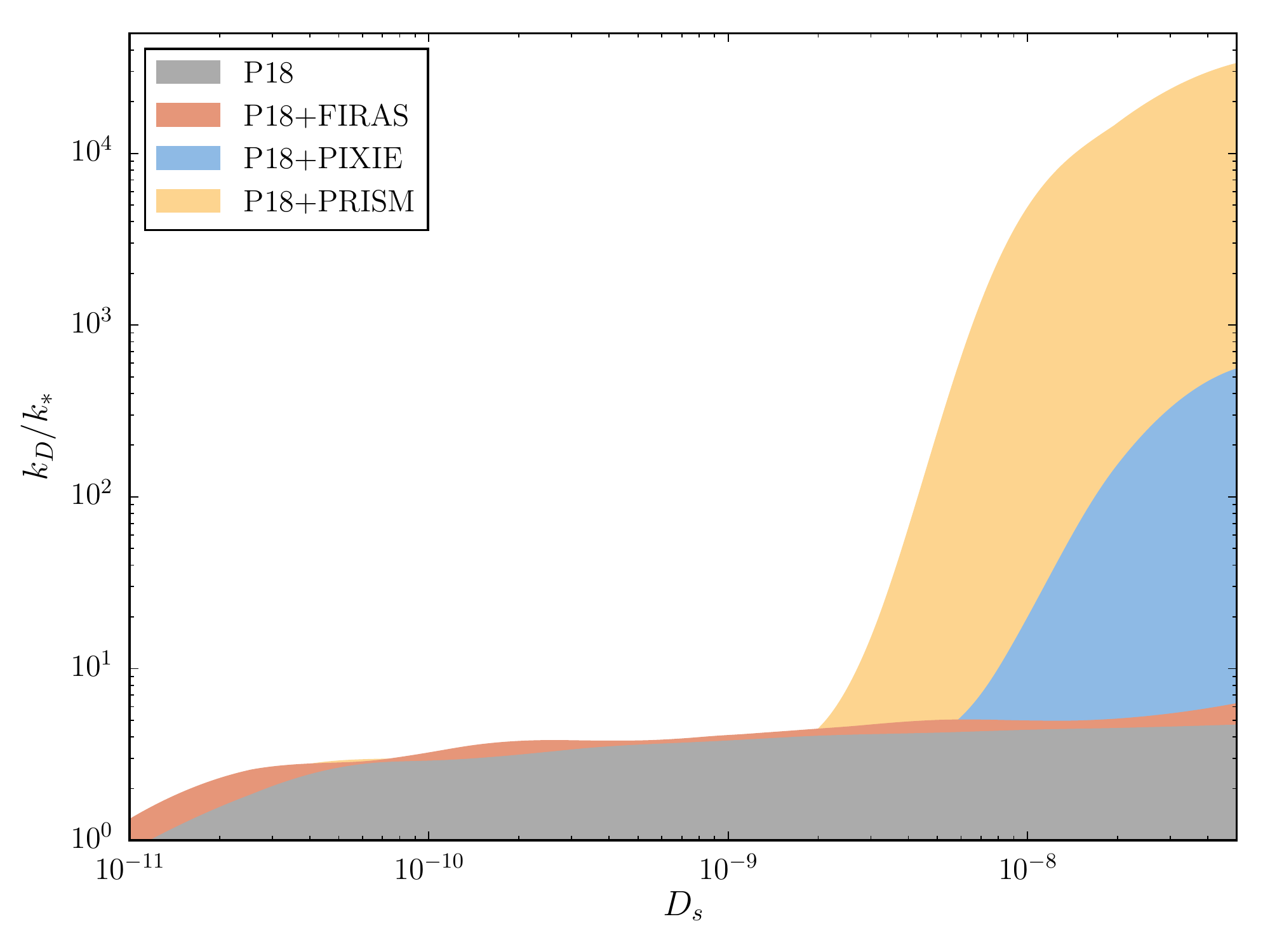}
	\includegraphics[width=7.5 cm]{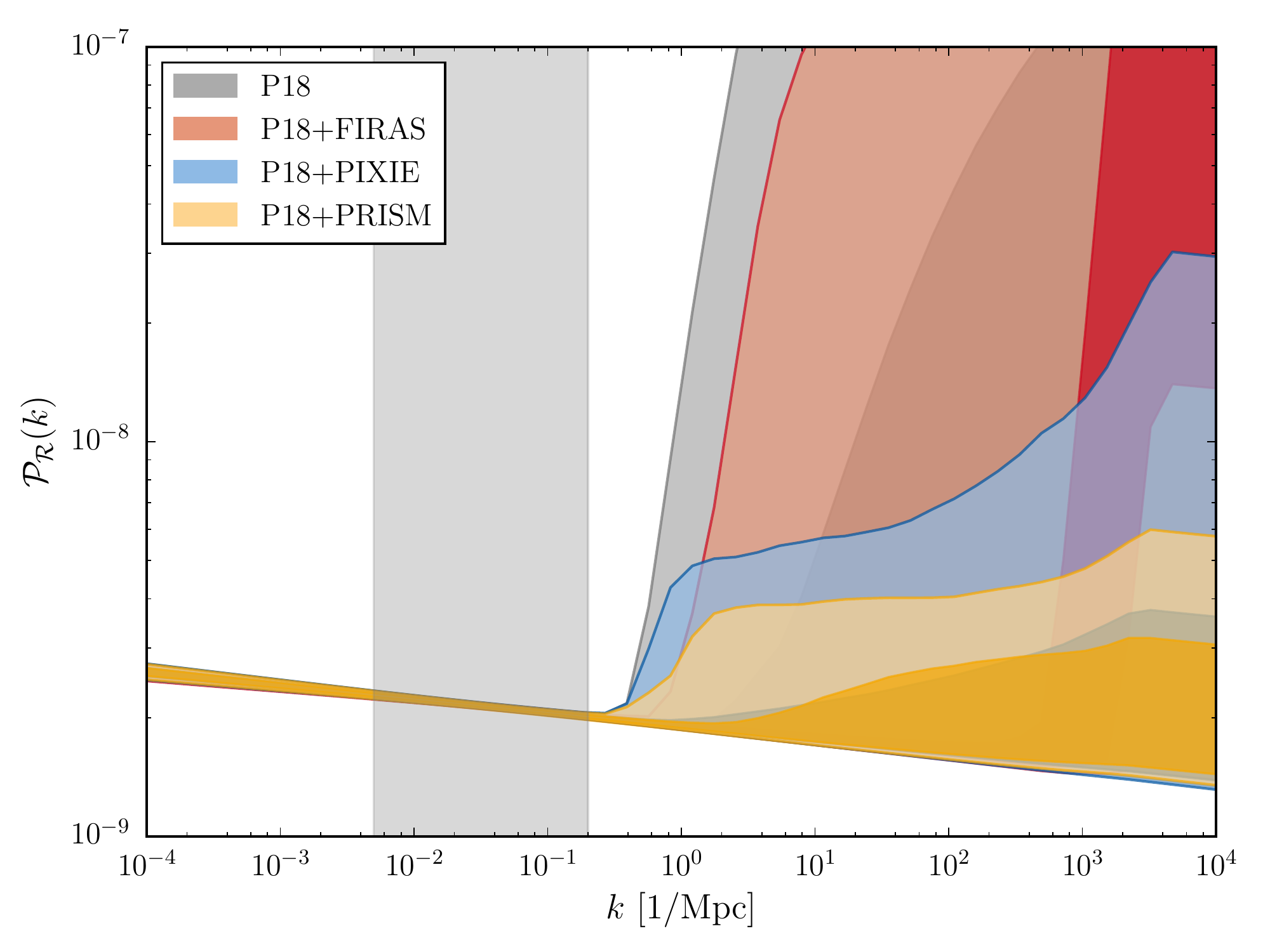}
	\caption{Same as in Fig. \ref{fig:p18_chluba}, but for the scenario with step-like features.}
	\label{fig:p18_steep}
\end{figure}

Finally, we investigate the constraints on various features in the PPS, focusing primarily on the possible presence of kinks and steps in the PPS, as well as of a multi-component inflationary scenario.
In these cases, we only show the results for the data set combinations involving Planck, since the considered features mainly play a role at Fourier modes above 1~Mpc$^{-1}$, as we will see. Therefore, the differences between Planck and the future combinations including CMB-S4+LiteBIRD are only secondary.
\dnew
First, we study a model with a broken power law, equivalent to the one already discussed in Sec. 3.3 of \cite{Chluba2012Inflaton}. We show in Fig.~\ref{fig:p18_chluba} a plot similar to Fig. 6 of the reference, but updated to include the latest Planck data constraints, a consistent marginalization over galactic and extra-galactic foregrounds as well as over the other cosmological parameters, and the contours for FIRAS and PRISM. The particular shape of the considered PPS involves a kink at a given scale $k_b$ (see Fig. 5 of \cite{Chluba2012Inflaton}), which can be formally parameterized as
\begin{align}
	\mathcal{P}_{\mathcal{R}}(k)=\left\{\begin{array}{ll}
	\mathcal{P}_{\mathcal{R}}(k) & \text { at } k<k_{\mathrm{b}} \\
	\mathcal{P}_{\mathcal{R}}\left(k_{\mathrm{b}}\right)\left(\frac{k}{ k_{\mathrm{b}}}\right)^{n_{\mathrm{s}}^{*}-1} & \text { at } k \geq k_{\mathrm{b}}
	\end{array}\right.\,,
\end{align} 
where $n_{\mathrm{s}}^{*}$ corresponds to the value of the scalar spectral index at scales above $k_b$. Such a power spectrum can be justified by several choices of inflationary potentials (for possible examples see e.g., the references given in~\cite{Chluba2012Inflaton}), and is often invoked as an example of a model with large perturbations on small scales.
\dnew
Although the general shape of the exclusion regions agrees well with the reference, the constraints displayed in the left panel of Fig. \ref{fig:p18_chluba} are significantly weakened compared to the reference, most notably due to the foreground and cosmological parameter marginalization performed in this work that was not included in \cite{Chluba2012Inflaton}. In order to explicitly show this, we investigate how much the constraints are strengthened when removing the marginalization of all foregrounds except for temperature shifts $\Delta_T$\, for a PIXIE-like experiment (resulting in a sensitivity of around $\delta \mu \sim 1.7 \cdot 10^{-7}$), and display the result as a black dotted line in the left panel of Fig. \ref{fig:p18_chluba}. As can be seen there, this degrades the constraints quite drastically, clearly showing the importance of the full marginalization for this purpose. 
However, note that this simplification of the marginalization procedure still does not fully reproduce the results presented in \cite{Chluba2012Inflaton}. 
In fact, the analysis technique employed in the reference is performed by imposing limits based on a cut of type $\mu < \delta \mu = 2 \cdot 10^{-7}$ (i.e., the nominal value reported in \cite{Kogut2011Primordial} and adopted by \cite{Chluba2012Inflaton}) instead of performing a likelihood analysis. 
Abandoning then the full treatment of the energy injection branching ratios used in \cite{Lucca2019Synergy} and instead using the prescription from \cite{Chluba2012Inflaton}, we obtain the solid black line shown in Fig. \ref{fig:p18_chluba}, which closely reproduces the constraints reported in the reference.
\dnew
Overall, as clear from the figure, we find that even up to scales of several thousands of Mpc$^{-1}$, SD experiments such as PIXIE and PRISM provide strong constraints on the possible values of the scalar tilt $n_s^{*}$ of the additional component, improving on the constraints obtained by the combination Planck+FIRAS. 
\dnew
Furthermore, in the right panel of Fig. \ref{fig:p18_chluba} we also show the corresponding reconstruction of the PPS for the given model at $1\sigma$ and $2\sigma$ (68.3\% CL and 95.4\% CL). 
The figure illustrates that the SD missions strongly impact the allowed deviation from a nearly scale invariant PPS up to very small scales (of around $10^4$~Mpc$^{-1}$), where the SD missions lose sensitivity due to the influence of the visibility function, as discussed in \cite{Fu:2020wkq}. 
Especially kinks towards higher $n_s^* > n_s$ are more constrained, as those significantly increase the heating rate, while kinks towards lower $n_s^* < n_s$ only negligibly reduce the heating rate. 
Note also that we imposed the constraint $k_b >10$ Mpc$^{-1}$, as done in \cite{Chluba2012Inflaton}, which is why our constraints only start deteriorating at larger wavenumbers.
\dnew
As the next example of possible feature in the PPS, we turn our attention to step-like features. In this case, we base our prescription on the analysis conducted in \cite{Byrnes2018Steepest} and, motivated by the results presented there, we limit the growth of the power spectrum to be at most proportional to $k^4$. Consequently, as a very rough approximation of Fig. 2 of \cite{Byrnes2018Steepest} (and in the same philosophy of Eqs. (15)-(18) of the reference), we choose to model the PPS as 
\begin{equation}
	\mathcal{P}_\mathcal{R}(k) = \begin{cases}
	A_s (k/k_*)^{n_s-1} & \text{, if $k<k_D$} \\ 
	A_s (k_D/k_*)^{n_s-1} (k/k_D)^4 & \text{, if $k_D<k<\widetilde{k}_D$} \\ 
	(A_s+D_s) (k/k_*)^{n_s-1} & \text{, if $\widetilde{k}_D < k$}
	\end{cases}~,
\end{equation}
where we have defined 
$\log\left(\,\widetilde{k}_D/k_D\right) = \log\left(\frac{A_s+D_s}{A_s}\right)/(5-n_s)$~,
ensuring the continuity of the PPS. Similar approaches have already been adopted in the literature (see e.g., Sec. 3.2 of \cite{Chluba2012Inflaton} and references therein), although again with the aforementioned limitations which we improve upon in this work.
\dnew
The resulting constraints on $k_D$ and $D_s$ are shown in the left panel of Fig. \ref{fig:p18_steep}. Although FIRAS does not improve on the constraints from Planck alone, both PIXIE and PRISM provide considerable improvements above scales of around $3k_* = 0.15$Mpc$^{-1}$, where Planck alone loses sensitivity completely. Notably, the inclusion of the SD mission PRISM allows to constrain the additional component $D_s$ to be approximately less than $10^{-8}$ for Fourier modes up to roughly $10^4$ Mpc$^{-1}$. This is nicely demonstrated in the right panel of Fig. \ref{fig:p18_steep}, where beyond the scales probed directly by Planck the constraints on the PPS immediately begin to weaken. The inclusion of SD missions strongly improves the Planck constraints in this region by many orders of magnitude.
\dnew
Finally, we focus on a model with two nearly scale invariant components of the PPS, which can be seen as generalization of models with continuous kinks at arbitrary scales, as in~\cite{Chluba2012Inflaton} (or in \cite{Clesse2014Testing}, where concrete models are also investigated, although no exclusion limits are derived there). These models involve two different scalar indices ($n_s$ and $m_s$) and amplitudes ($A_s$ and $B_s$), and we require as a prior for the additional parameters that $B_s>0$ and $m_s>1$, such that the additional component can never be perfectly degenerate with the standard PPS parameters, and is forced to be flat or growing with the wavenumber.

\begin{figure}[t]
	\centering 
	\includegraphics[width=7.5 cm]{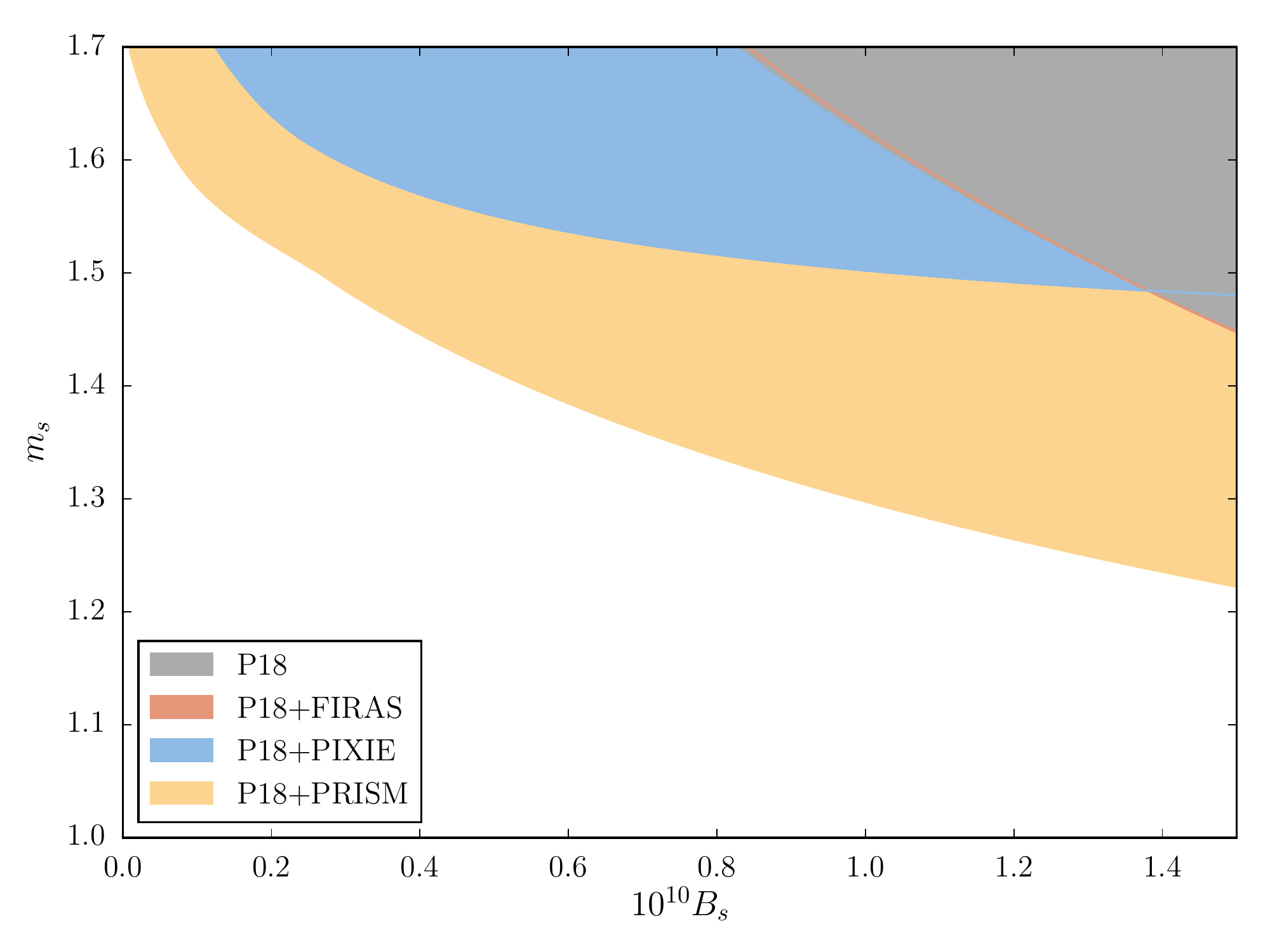}
	\includegraphics[width=7.5 cm]{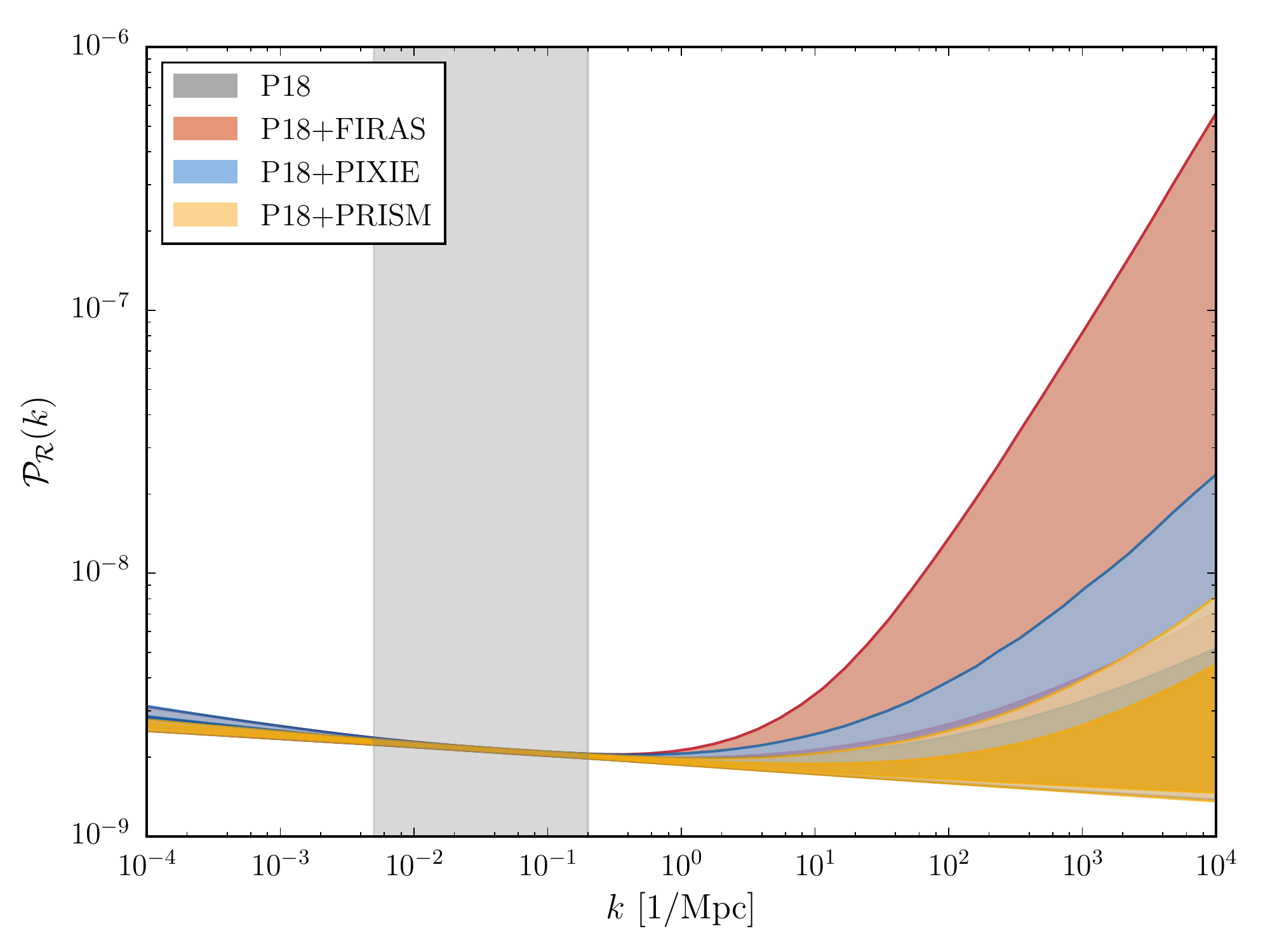}
	\caption{Same as in Fig. \ref{fig:p18_chluba}, but for the scenario with two components.}
	\label{fig:p18_twocomp}
\end{figure}

\dnew
We show the bounds on the parameters of the additional component in the left panel of Fig. \ref{fig:p18_twocomp}. Similarly to the previous cases, although FIRAS does not improve the bounds from Planck alone, PIXIE and especially PRISM can provide much stronger constraints. Indeed, the inclusion of the latter SD missions allow to constrain significantly smaller amplitudes $B_s$ of the additional component, improving on the Planck alone case by around one order of magnitude. The constraints on the slope $m_s$ are also significantly tighter when including SD missions such as PRISM, although they become increasingly weaker as $B_s \to 0$. Furthermore, in the right panel of Fig. \ref{fig:p18_twocomp} we can observe that the constraints on the PPS behave similarly to the previous cases for Fourier modes above 1 Mpc$^{-1}$, with the difference that here strong deviations in the intermediate range of $1-100$Mpc$^{-1}$ are excluded. This is a natural consequence of the constant tilt of the additional component\footnote{To achieve a strong deviation on scales of $1-100$Mpc$^{-1}$ a high $m_s$ would be required and since $m_s$ is constant one would expect an even stronger deviation at scales of $10^3-10^4$Mpc$^{-1}$ by many orders of magnitude.} and directly translates into the inverse proportionality of the bound between $m_s$ and $B_s$ mentioned before.

\section{Conclusions}\label{sec:con}

There is a well known synergy between CMB anisotropies and SDs, with far-reaching applications ranging from inflation to beyond the standard model physics. In this paper we focus our attention on its possible applications to inflationary models that go beyond the simple addition of a running of the scalar spectral index, and indeed even beyond the common assumption of slow-roll inflation. In particular, we investigate higher-order polynomial expansions of the inflationary potential as well as a slow-roll expansion of the PPS, models with kinks and steps in the PPS, and a two-component model for the PPS. All of these extensions are theoretically motivated by various different scenarios, including primordial GWs and PBHs, and could offer insight into these interesting phenomena.
\dnew
In order to extensively explore this interplay between CMB anisotropy and SD missions over more than eight orders of magnitude in Fourier space for such a variety of inflationary scenarios, we consider several combinations of concluded, upcoming, and proposed missions such as Planck, CMB-S4 and LiteBIRD, as well as FIRAS, PIXIE, and PRISM. Furthermore, building on the work already presented in \cite{Lucca2019Synergy, Fu:2020wkq}, in our analyses we include a complete implementation of the many galactic and extra-galactic foregrounds which arise from reionization and late-time sources and that are very relevant for the considered SD missions. This addition provides a more realistic treatment of the employed SD likelihoods and increases the solidity of the discussed results. As such, this work presents the first constraints on inflationary models from CMB SDs taking into account the marginalization over foregrounds.
\dnew
For the scenarios involving a polynomial expansion of the of the inflationary potential there are mainly two ways in which SDs can improve on CMB anisotropy measurements alone. First, they allow to increase the lever-arm over which the PPS is constrained by about four orders of magnitude in $k$-space. Second, the addition of a SD mission allows to extend the range of observable Fourier modes. As a consequence, the observable window of inflation spans a larger number of $e$-folds, and the mere fact requiring the Taylor-expanded potential to support a longer inflationary epoch sharply limits the freedom of high-order terms. Therefore, thanks to this second effect, any SD mission, even with FIRAS-like sensitivities, is intrinsically able to considerably improve on CMB anisotropy constraints alone, simply because of the extended observable scales.
\dnew
Concretely, considering an expansion up to fourth order of the inflationary potential, we find that SDs could improve the constraints on higher-order terms by up to 350\% with respect to Planck alone and 70\% with respect to the future combination of \text{CMB-S4+LiteBIRD}. Moreover, for an expansion up to sixth order we find that the inclusion of SDs would lead to tighter constraints by up to approximately 640\% and 630\% compared to Planck and \text{CMB-S4+LiteBIRD} alone, respectively. This is because the  higher-order terms are much more significantly constrained by the lengthened lever-arm, while they barely benefit from the increased precision on CMB scales. Therefore, future SD mission promise to significantly aid in PPS reconstruction analyses based on the inflationary potential.
\dnew
Assuming instead that the logarithm of the PPS can be described by a Taylor expansion in $\ln k$ with coefficients that can be related to the inflaton potential through the slow-roll formalism, our results are similar to those of the previous scenarios. In fact, also in this case future SD missions would provide an improvement on the parameter constraints of up to 200\% together with Planck, and up to 180\% when combined with \text{CMB-S4+LiteBIRD}. However, the average improvement of the parameter uncertainty is significantly reduced in the \text{CMB-S4+LiteBIRD} case. 
\dnew
Additionally, we also consider a variety of features that could affect the shape of the PPS. First, we focus on the possible presence of kinks in the PPS, showing that the addition of SDs would provide strong constraints on the scalar tilt of the additional component up to scales of roughly 10$^{4}$ Mpc$^{-1}$. Furthermore, we also allow for steps in the PPS. As a result, we obtain that a SD mission such as PRISM would constrain the amplitude of the additional component to be $B_s < 10^{-8}$ for $k$ values below $10^4$ Mpc$^{-1}$, whereas Planck alone can only place bounds below approximately $10$ Mpc$^{-1}$. Finally, we also investigate the possibility of having a second component of the PPS in addition to the standard one with the same functional form. Also in this case we observe great improvements from the addition of SDs. For all of these cases we show by means of a reconstruction of the PPS in each specific model that SD missions would be able to put tight bounds on the PPS on scales much larger than around 1 Mpc$^{-1}$\,.
\dnew
In conclusion, the results presented here clearly show that the future observation of SDs of the CMB would not only extend our understanding of inflation in terms of observable scales, but it would also help test the validity of the assumption of scale invariance with unprecedented precision, nicely complementing current and future CMB anisotropy measurements.

\section*{Acknowledgements}
We thank Julien Lesgourgues for the  insightful discussions that kickstarted this project. We also thank Julien Lesgourgues and Jens Chluba for very useful feedback on the manuscript.
NS acknowledges support from the DFG grant \mbox{LE~3742/4-1}. 
ML is supported by an F.R.S.-FNRS fellowship, by the \tquote{Probing dark  matter with neutrinos} ULB-ARC convention and by the IISN convention 4.4503.15.
DH is supported by the FNRS research grant number \mbox{F.4520.19}.   
Simulations for this work were performed with computing resources granted by JARA-HPC from RWTH Aachen University under the project jara0184.

\appendix 
\section{From the primordial power spectrum to the CMB}\label{app:th_CMB}

In this Appendix we will briefly review the connection between the shape of the PPS and that of the CMB anisotropy angular power spectra. For conventions and notations, we refer to the pedagogical overview presented in Sec. 3 of \cite{Hooper:2019xer} (see also the many references therein for related discussions).
\dnew
In the case of CMB anisotropies, there is a direct dependence of the different harmonic power spectra $C_\ell$ and the shape of the PPS. For instance, in the case of the temperature anisotropy spectrum, it is possible to expand the relative temperature differences $\delta T/T$ in a given direction $\hat{n}$ in spherical harmonics as
\begin{align}
	\frac{\delta T}{T}(\hat{n})=\sum_{\ell m} a_{\ell m} Y_{\ell m}(\hat{n})\,,
\end{align}
where $ Y_{\ell m}(\hat{n})$ are Legendre polynomials and $a_{\ell m}$ the corresponding coefficients that can be shown to have the form
\begin{align}\label{eq:a_lm}
	a_{\ell m}=(-i)^{\ell} \int \frac{d^{3} k}{2 \pi^{2}} Y_{\ell m}(\hat{k}) \Theta_{T, \ell}\left(\tau_{0}, \textbf{k}\right)\,,
\end{align}
depending on the Fourier modes $\Theta_{T, \ell}\left(\tau_{0}, \textbf{k}\right)$ at a given time $\tau_{0}$. In the last equation we have implicitly transformed the integral from polar coordinates to Fourier space. The correlation function between the coefficients
\begin{align}\label{eq:Cl}
	\left\langle a_{\ell m} a_{\ell m}^{*}\right\rangle =\frac{1}{2 \pi^{2}} \int \frac{d k}{k}\left[\Theta_{T, \ell}\left(\tau_{0}, k\right)\right]^{2} \mathcal{P}_\mathcal{R}(k)\equiv C_{\ell}
\end{align}
provides then a theoretical prediction that can be compared to observations. Here, $\Theta_{T, \ell}\left(\tau_{0}, k\right)$ is the harmonic transfer function of the photon temperature perturbations, and $\mathcal{P}_\mathcal{R}(k)$ can be computed according to \fulleqref{eq:dim_pps}. These two quantities are related to the Fourier modes introduced in \fulleqref{eq:a_lm} according to
\begin{align}
	\left\langle\Theta_{T, \ell}(\tau, \textbf{k}) \Theta_{T, \ell}^{*}(\tau, \textbf{k}')\right\rangle=\frac{2 \pi^{2}}{k^{3}} \mathcal{P}_\mathcal{R}(k)\left[\Theta_{T, \ell}(\tau, k)\right]^{2} \delta^{(3)}\left(\textbf{k}-\textbf{k}'\right)\,.
\end{align}
As clear from \fulleqref{eq:Cl}, the PPS amplitude $A_s$ determines the overall amplitude of the temperature power spectrum and the spectral index $n_s$ its tilt. 
\dnew
It is then possible to compare these theoretical predictions to the latest observational evidence gathered by the Planck satellite\footnote{Including information from the measurement of BAO affects only marginally these values \cite{Akrami2018PlanckX}.} to obtain the best-fitting values for the PPS parameters $\ln \left(10^{10} A_{s}\right) = 3.049 \pm 0.015$,  $n_{\mathrm{s}} = 0.9625 \pm 0.0048$, $\alpha_s = 0.002 \pm 0.010$,  ${\beta_s = 0.010 \pm 0.013}$ (see Tab. 5 of \cite{Aghanim2018PlanckVI} and Eqs. (16) to (18) of \cite{Akrami2018PlanckX}). As nicely represented e.g. in Fig. 22 of \cite{Akrami2018PlanckX}, these values are based on the measurement of $\ell$ values between 2 and approximately $2 \cdot 10^3$, which correspond to Fourier modes $k$ roughly between $10^{-4}$ Mpc$^{-1}$ and 0.15 Mpc$^{-1}$.

\newpage

\bibliography{bibliography}
\bibliographystyle{JHEP}
\end{document}